\documentclass[10pt, conference]{IEEEtran}

\makeatletter
\def\ps@headings{%
\def\@oddhead{\mbox{}\scriptsize\rightmark \hfil \thepage}%
\def\@evenhead{\scriptsize\thepage \hfil \leftmark\mbox{}}%
\def\@oddfoot{}%
\def\@evenfoot{}}
\makeatother
\ifCLASSINFOpdf
\else
\fi

\usepackage{lscape, latexsym, amssymb, amsmath, multirow}
\usepackage[linesnumbered, vlined, ruled]{algorithm2e}
\usepackage[pdftex]{graphicx}
\usepackage{graphicx}
\usepackage{subfigure}
\usepackage{multirow}
\usepackage{mathtools, bbm, color}
\usepackage{pdfpages}

\usepackage{amsmath,amssymb,amsthm,mathrsfs,amsfonts,dsfont}

\usepackage{pifont, textcomp, marvosym, wasysym, extarrows}
\usepackage{tikz}

\usepackage{algpseudocode}

\newcommand{\tabincell}[2]{\begin{tabular}{@{}#1@{}}#2\end{tabular}}

\newtheorem{theorem}{Theorem}

\newtheorem{corollary}{Corollary}

\begin{document}

\title{De-Health: All Your Online Health Information Are Belong to Us}

\author{
Shouling Ji$^{\dag, \ddag}$, Qinchen Gu$^\ddag$, Haiqin Weng$^\dag$, Qianjun Liu$^\dag$, 
Qinming He$^\dag$, Raheem Beyah$^\ddag$, Ting Wang$^\sharp$ \\

$^\dag$ Zhejiang University \\
$^\ddag$ Georgia Institute of Technology\\
$^\sharp$ Lehigh University \\

sji@zju.edu.cn
}
\maketitle

\begin{abstract}
In this paper, we study the privacy of online health data.
We present a novel online health data De-Anonymization (DA) framework,
named \emph{De-Health}. De-Health consists of two phases:
\emph{Top-$K$ DA}, which identifies a candidate set
for each anonymized user, and \emph{refined DA},
which de-anonymizes an anonymized user to a user in its candidate set.
By employing both candidate selection and DA verification schemes,
De-Health significantly reduces the DA space by several orders of magnitude
while achieving promising DA accuracy.
Leveraging two real world online health datasets WebMD (89,393 users, 506K posts)
and HealthBoards (388,398 users, 4.7M posts), we validate the efficacy
of De-Health. Further, when the training data are insufficient,
De-Health can still successfully de-anonymize a large portion of anonymized users.

We develop the first analytical framework on the soundness and effectiveness
of online health data DA. By analyzing the impact
of various data features on the anonymity, we derive the conditions and probabilities
for successfully de-anonymizing one user or a group of users
in exact DA and Top-$K$ DA.
Our analysis is meaningful to both researchers and policy makers
in facilitating the development of more effective anonymization techniques
and proper privacy polices.

We present a linkage attack framework which can link online health/medical
information to real world people. Through a proof-of-concept attack,
we link 347 out of 2805 WebMD users to real world people, and find
the
full names, medical/health information, birthdates, phone numbers, and
other sensitive information for most of the re-identified users.
This clearly illustrates the fragility of the notion of privacy of those who use online health forums.
\end{abstract}

%

%
\IEEEpeerreviewmaketitle

\section{Introduction} \label{intro}

\textbf{Status Quo.}
The advance of information technologies has greatly transformed
the delivery means of healthcare services: from traditional
hospitals/clinics to various online healthcare services.
This fact can be well explained by Charles Simmons, a software engineer in Los Angeles, California \cite{hbwiki}.
\begin{quote}
\emph{In 1997, after experiencing a variety of symptoms for which doctors
had no explanation, Simmons turned to the Web for answers and support.
When he did not find online support groups in the areas he needed,
he realized that there was a need for a health support website covering
a wide range of health topics.}
\end{quote}
Ever since their introduction, online health services experienced rapid growth,
and have had millions of users and accumulated billions of users' medical/health records
\cite{webmdreport}\cite{hb}.

According to several national surveys,
$\sim 59\%$ of US adults have employed the Internet (online health information)
as a diagnostic tool in 2012 \cite{survey1}, and
on average, the US consumers spend $\sim 52$ hours annually to search for and peruse
online health information while only visiting doctors three times
per year in 2013 \cite{survey2}. Moreover, ``on an average day,
$6\%$ of the US Internet users perform online medical searches to better prepare for
doctors' appointments and to better digest information obtained from
doctors afterwards" \cite{survey3}. Therefore, online health services
play a more important role in people's daily life.

When serving users (we use patients and users interchangeably in this paper),
the online health services accumulate a huge amount of the users' health data.
For instance, as one of the leading American corporations
that provide health news, advice, and expertise \cite{webmd}\cite{webmdwiki},
WebMD reached an average of approximately 183 million monthly unique visitors and
delivered approximately 14.25 billion page views in 2014 \cite{webmdreport}.
Another leading health service provider, HealthBoards (HB),
has over 10 million monthly visitors, 850,000 registered members,
and over 4.5 million health-related/medical messages posted \cite{hb}.
Due to the high value of enabling low-cost, large-scale data mining
and analytics tasks, e.g., disease transmission and control research \cite{tilhoupnas10},
disease inference \cite{niewantkde15},
and predicting future instances of domestic abuse \cite{reikohbmj09},
those user-generated health data are increasingly
shared, disseminated, and published for research \cite{gkoloujbi14},
business \cite{webmdwiki}\cite{hbwiki}, government applications \cite{usgov}\cite{cagov},
and other scenarios \cite{gkoloujbi14}\cite{hribloamia14}.

\textbf{Privacy Issues of Online Health Data.}
In addition to the high value for various applications,
online health data carry numerous sensitive details of the users that
generate them \cite{gkoloujbi14}\cite{hribloamia14}\cite{emarodbmj15}.
Therefore, before sharing, disseminating, and publishing the health data,
proper privacy protection mechanisms
should be applied and privacy policies
should be followed.
However, the reality is that it is still an open problem for protecting
online health data's privacy with respect to both the \emph{technical} perspective
and the \emph{policy} perspective.

From the technical perspective, most existing health data anonymization
techniques (which are called \emph{de-identification} techniques in the
medical and policy literature \cite{gkoloujbi14}\cite{emarodbmj15}), e.g., the techniques in
\cite{emaarbjmir12}-\cite{gkoloujbi14}, if not all,
focus on protecting the privacy of \emph{structured medical/health data}
that are usually generated from hospitals, clinics, and/or other
official medical agencies (e.g., labs, government agencies).
Nevertheless, putting aside their performance and effectiveness,
existing privacy protection techniques for structured health data
can hardly  be applied to online health data due to the following
reasons \cite{niezhatkde14}\cite{luotanir08}\cite{niewantkde15}.
($i$) \emph{Structure and heterogeneity:} the structured health data
are well organized with structured fields while online health data
are usually heterogeneous and structurally complex.
($ii$) \emph{Scale:} a structured health dataset usually consists
of the records of tens of users to thousands of users \cite{emaarbjmir12}-\cite{gkoloujbi14},
while an online health dataset can contain millions of users
\cite{webmdreport}\cite{hb}\cite{niewantkde15}.
($iii$) \emph{Threat:} Compared to online health data, the dissemination of structured
health data is usually easier to control, and thus the
potential of a privacy compromise is less likely. Due to its open-to-public nature, however,
the online health data dissemination is difficult to control,
and adversaries may employ multiple kinds of means and auxiliary information
to compromise the data's privacy (we will show this later in this paper).

From the policy making perspective,
taking the US Health Insurance Portability and Accountability Act (HIPAA) \cite{hippa} as an example,
although HIPAA sets forth methodologies for anonymizing health data
(including online health data), once the data are anonymized,
they are no longer subject to HIPAA regulations and can be used for any purpose.
However, when anonymizing the data, HIPAA does not specify any concrete technique
other than high-level guidelines. Therefore, naive anonymization technique may
be applied.

\textbf{Our Work.}
Toward helping users, researchers, data owners, and policy makers
comprehensively understand the privacy vulnerability of online health data,
we study the privacy of online health data.
Specifically, we focus on the health data generated on
online health forums like WebMD \cite{webmd} and HB \cite{hb} in this paper.
These forums disseminate personalized health information
and provide a community-based platform for connecting patients
with doctors and other patients via interactive question and answering, symptom analysis,
medication advice and side effect warning, and other interactions
\cite{webmd}\cite{hb}\cite{niezhatkde14}.

As we mentioned earlier, a significant amount of medical records
have been accumulated in the repositories of these health
websites. According to the website privacy policies \cite{webmd}\cite{hb},
they explicitly state that they collect personal information of users (patients),
including contact information, payment information,
geographic location information, personal profile, medical information,
transaction information, Cookies, and other sensitive information.
For instance, in WebMD's privacy policy \cite{webmd}, they state that
\begin{quote}
\emph{We may collect ``Personal Information" about you --
such as your name, address, telephone number, email address or health information $\cdots$
We may collect ``Non-Personal Information" -- information that cannot
be used by us to identify you -- via Cookies, Web Beacons,
WebMD mobile device applications and from external sources,
even if you have not registered with or provided any Personal Information to WebMD $\cdots$}
\end{quote}
and similarly, in HB' privacy policy \cite{hb},
\begin{quote}
\emph{We collect personal information for various business purposes
when you interact with us $\cdots$
We collect information about you in two basic ways:
First, we receive information directly from you.
Second, through use of cookies and other technologies,
we keep track of your interactions $\cdots$}
\end{quote}
To use those online health services, users have to accept
their privacy policies. For instance, in HB' privacy policy,
it is explicitly indicated that ``if you do not agree to this privacy policy,
please do not use our sites or services".
Therefore, using the online health services requires the enabling of
those service providers like WebMD and HB to collect
users' personal informaiton.

As stated, the collected personal information will be used
for research and \emph{various business
purposes}, e.g., data mining tasks and precise advertisements from
pharmaceutical companies.
Although these medical records are only affiliated with
some user-chosen pseudonyms or anonymized IDs, some natural questions arise:
when those data are shared with commercial partners (one of the most typical
various business purposes)\footnote{The business model (revenue model) of
most online health forums is advertisement based \cite{webmdwiki}\cite{hb}.},
or published for research,
or collected by adversaries, can they be de-anonymized even if the patients
who generated them are anonymized? and can those medical records be connected
to real world people? In this paper, we answer these two questions
by ($i$) proposing a novel online health data De-Anonymization (DA) framework;
($ii$) providing a general theoretical analysis
for the soundness and effectiveness of online health data DA;
and ($iii$) discussing how to link the medical records
to real world people.
We also discuss the implications of our findings
to online health data privacy researchers, users, as well as policy makers.

\textbf{Our Contributions.} Our key contributions are the following:

  (1)
  We present a novel DA framework, named \emph{De-Health}, for large-scale
  online health data. De-Health is a two-phase DA framework.
  In the first phase, De-Health performs \emph{Top-$K$ DA}.
  It first constructs a \emph{User-Data-Attribute (UDA) graph}
  based on the data correlation among users, and then identifies structural features
  (i.e., graph features) from the UDA graph.
  Leveraging those structural features, a \emph{Top-$K$ candidate set} is constructed
  for each anonymized user.
  In the second phase, \emph{refined DA} is performed.
  Leveraging both correlation and stylometric features
  of an anonymized user and the users in her corresponding Top-$K$ candidate set,
  De-Health trains a classifier using benchmark machine learning techniques
  to de-anonymize the anonymized user
  to some user in its candidate set.
  De-Health has two distinguishing features: ($i$) by utilizing the UDA graph
  and Top-$K$ candidate sets, it can be easily scaled to large-scale health data
  with high accuracy preservation;
  and ($ii$) De-Health can be applied to both \emph{closed-world} DA
  (each anonymized user appears in the training/auxiliary data)
  and \emph{open-world} DA
  (there are some anonymized users that may not appear in the training/auxiliary data).

  (2)
  We provide a general theoretical analysis framework for the soundness
  and effectiveness of online health data DA.
  In the framework, we analyze the impacts of structural features
  and stylometric features on the anonymity of health data.
  Specifically,
  we quantify the conditions and probabilities
  of successfully de-anonymizing (including exact DA
  and Top-$K$ DA) one user or a group of users.
  The theoretical analysis has meaningful implications to health data
  privacy research  and policy making:
  understanding the impacts of features on the data's anonymity
  will facilitate researchers and policy makers to develop more effective anonymization techniques
  and proper privacy policies.

  (3)
  Leveraging two real world online health datasets WebMD (89,393 users, 506K posts)
and HB (388,398 users, 4.7M posts), we conduct extensive
  evaluations to examine the performance of De-Health in closed-world
  and open-world DA settings.
  The results show that the Top-$K$ DA of De-Health is very
  powerful on large-scale datasets. By seeking a
  Top-$K$ candidate set for each anonymized user,
  the DA space is effectively decreased by
  several orders of magnitude while
  providing high accuracy preservation, which enables the development
  of an elegant machine learning based classifier for refined DA.
  Even when little data are available for training,
  De-Health can still achieve a satisfying DA performance,
  which significantly outperforms the traditional DA approach.

  (4)
  We present a linkage attack framework,
  which can link online health service users to other Internet services
  as well as real world people.
  We validate the framework leveraging proof-of-concept attacks.
  For instance, it can successfully
  link 347 out of 2805 (i.e., $12.4\%$) target WebMD users to
  real world people, and find most of their full names, medical/health information,
  birthdates, phone numbers, addresses, and other sensitive information.
  Thus, those users' privacy can be compromised and one can learn
  the sexual orientation and related infectious diseases, mental/psychologicla problems,
  and suicidal tendency from some users' health/medial data.

\section{Data Collection \& Feature Extraction} \label{datacollection}

\subsection{Data Collection}

We collect online medical postings from two leading US
online health services providers WebMD \cite{webmd} and HB \cite{hb}.
As the leading health portal in the US \cite{webmdwiki},
WebMD provides valuable health information and tools for managing its users'
health, and support to those who seek online health information.
HB provides a one-stop support group community offering more than
200 message boards on various diseases, conditions, and health topics.
It was rated as one of the top 20 health information websites by Consumer Reports
Health WebWatch \cite{hb}\cite{hbwiki}.
We collected the health data of WebMD and HB registered users
for approximately $4$ months, from May to August, 2015.
This collection process resulted in 540,183 webpages from WebMD and
2,596,433 webpages from HB. After careful analysis and processing,
we extracted 506,370 disease/condition/medcine posts that were generated by
89,393 registered users from the WebMD dataset (5.66 posts/user on average) and
4,682,281 posts that were generated by 388,398 registered users
from the HB dataset (12.06 posts/user on average).
We show some example posts from WebMD and HB in Appendix \ref{posts}.
From the example posts, we can see that a lot of sensitive information
of the users can be learned.

\begin{figure}
\centering
\includegraphics[width=2in]{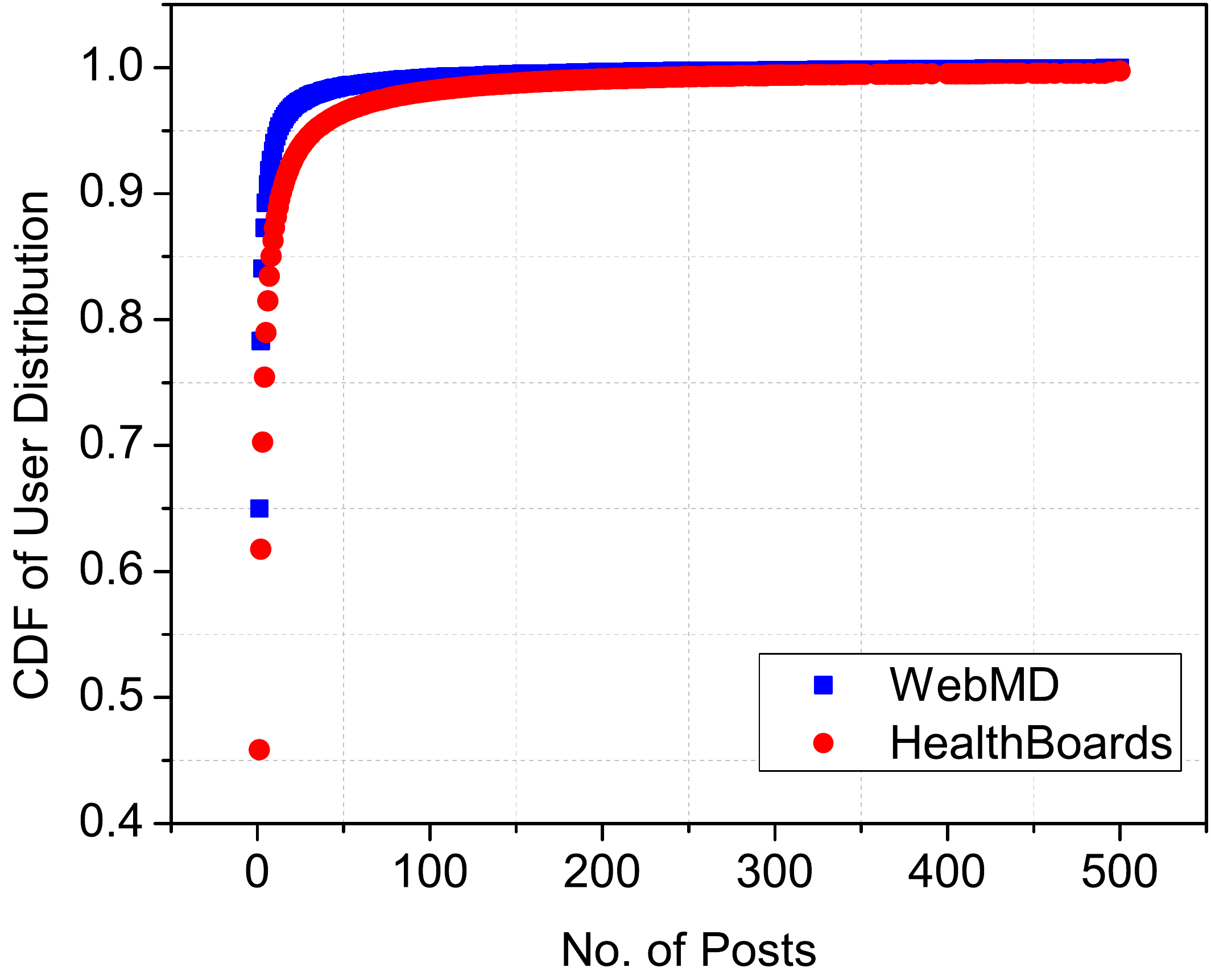}
\caption{CDF of users with respect to the number of posts.} \label{f_postdis}
\end{figure}

\begin{figure}
\centering
\includegraphics[width=2in]{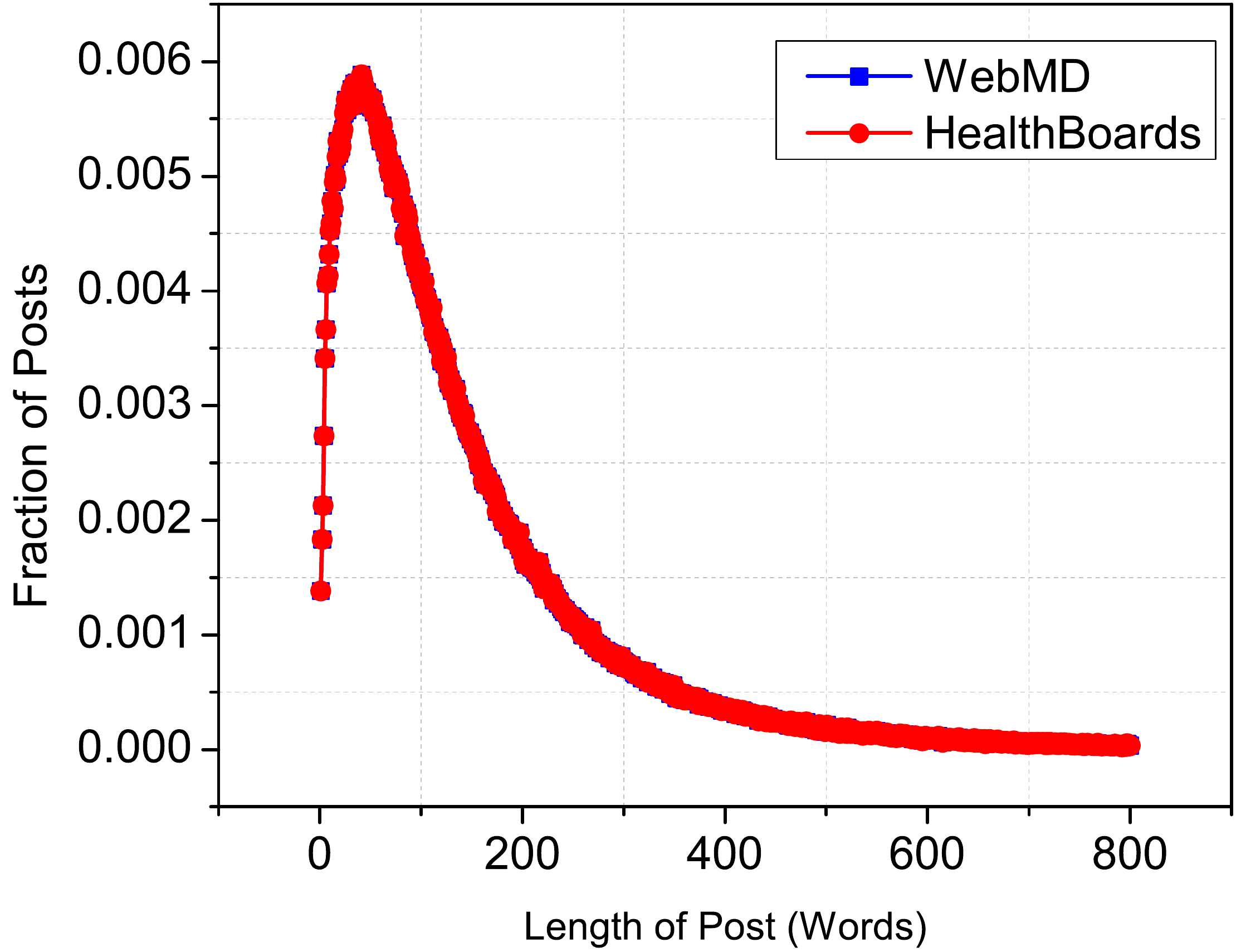}
\caption{Post length distribution.} \label{f_postlength}
\end{figure}

We show the Cumulative Distribution Function (CDF) of the
number of users with respect to the number of posts in Fig.\ref{f_postdis},
from which we observe that most of the users only have a few
posts, e.g., $87.3\%$ WebMD users and $75.4\%$ HB users
have less than 5 posts. We further show the length distribution
of the posts in WebMD and HB in terms of the number of words
in Fig.\ref{f_postlength}. Most of the posts in the two datasets
have a length less than 300 words. On average, the length of WebMD posts
is 127.59 and the length of HB posts is 147.24.

\subsection{User-Data-Attribute Graph \& Feature Extraction}

\textbf{User Correlation Graph.}
Online health services provide a platform for connecting patients
via interactive disease and symptom discussion, health question answering,
medicine and possible side effect advice, etc.
For instance, on WebMD and HB, when one disease topic is raised
by some user, other users may join the discussion of this topic by providing
suggestions, sharing experience, making comments, etc.
Due to this fact, HB is also classified as a
health-oriented social networking service \cite{hbwiki}.

Therefore, if we take such user interactivity into consideration,
there is some correlation, i.e., the co-disease/health/medicine discussion relation,
among users. To characterize such interactivity,
we construct a \emph{user correlation graph} based on the
relationships among the data (posts) of different users.
Particularly, for each user in the WebMD/HB dataset,
we represent him/her as a node in the correlation graph.
For two users $i$ and $j$, if they post under the same
health/disease topic, i.e., they made posts on the same topic
initialized by some user (could be $i$, $j$, or some other user),
we consider that there is an undirected edge, denoted by $e_{ij}$, between $i$ and $j$.
Furthermore, we note that the number of interactive discussions
between different pairs of users might be different.
Therefore, we assign a \emph{weight} for each edge to characterize
the interactivity strength, which is defined as the number of times
that the corresponding two users co-discussed under the same topic.

Now, we formally define the user correlation graph as $G = (V, E, W)$,
where $V = \{1, 2, \cdots, n\}$ denotes the set of users,
$E = \{e_{ij} | i, j \in V\}$ denotes the set of edges among users,
and $W = \{w_{ij} | $ for $e_{ij} \in E, w_{ij}$ is the weight (interactivity strength)
associated with edge $e_{ij}\}$.
For $i \in V$, we define its \emph{neighborhood} as $N_i = \{j | e_{ij} \in E\}$.
Let $d_i = |N_i|$ be the number of neighbor users of $i$, i.e., the \emph{degree} of user $i$.
When taking the weight information into consideration,
we define $wd_i = \sum\limits_{j \in N_i} w_{ij}$ to be the \emph{weighted degree} of $i$.
For our following application, we also define a \emph{Neighborhood Correlation Strength}
(NCS) vector for each user. Specifically, for $i \in V$, its NCS vector
is defined as $\mathbf{D}_i = <w_{ij}' | j \in N_i>$, where
$<w_{ij}' | j \in N_i>$ is a decreasing order sequence of
$\{w_{ij} | j \in N_i\}$.
Given $i, j \in V$,
we define the \emph{distance} (resp., \emph{weighted distance}) between $i$ and $j$ as the length
of the shortest path from $i$ to $j$ in $G$ when the weight information is overlooked
(resp., considered), denoted by $h_{ij}$ (resp., $wh_{ij}$).

We analyze the degree distributions of the WebMD graph and
the HB graph, as well as the community structure of the
WebMD graph in Appendix \ref{degcom}.
Basically the graph's connectivity is not strong (average degree is low and
graph is not connected).

\textbf{Stylometric Features.}
Using writing style for author attribution
can be traced back to the 19th century \cite{menscience87}.
Recently, stylometric approaches have been applied to broad security and
privacy issues, from author attribution \cite{abbchetis08}\cite{narpassp12}\cite{stoovewg14}
to fraud and deception detection \cite{afrbresp12},
underground cybercriminal analysis \cite{afrcalsp14}, and programmer DA \cite{calharusenix15}.
According to the findings in those applications,
users have distinctive writing styles (especially, in the non-adversarial scenario).
Thus, when providing sufficient data (written materials, e.g., blogs, documents, passages),
many users can be uniquely identified/de-anonymized from a (large) group of
candidate users using benchmark
machine learning models trained by their stylometric features \cite{abbchetis08}\cite{kopschlre11}\cite{narpassp12}\cite{stoovewg14}.
Furthermore, as demonstrated in \cite{afrbresp12}, it is difficult for users
to intentionally obfuscate their writing style or attempt to
imitate the writing styles of others in a long term.
Moreover, even that happens, with a high probability, specific linguistic features can still be
extracted from the long term written materials to identify the users.
Therefore, for our purpose, we seek to employ the linguistic features of
the health data (posts written by users) to de-anonymize the associated users.

\begin{center}
\begin{table}[!tp]
 \caption{Stylometric features.} \label{tab_feature}
  \centering
  \begin{tabular}{ l  l  l}
    \hline

    \textbf{Category} & \textbf{Description}  & \textbf{Count} \\
    \hline
    \hline
    Length & \tabincell{l}{\# of characters and paragraphs, \\average
    \# of characters per word} & 3 \\
    \hline
    Word Length & \tabincell{l}{freq. of words of different lengths} & 20 \\
    \hline
    \tabincell{l}{Vocabulary \\richness} & \tabincell{l}{Yule's K, hapax /tris/dis/tetrakis \\legomena} & 5 \\
    \hline
    Letter freq. & \tabincell{l}{freq. of `a/A' $\sim$ `z/Z'} & 26 \\
    \hline
    Digit freq. & \tabincell{l}{freq. of `0' $\sim$ `9'} & 10 \\
    \hline
    \tabincell{l}{Uppercase letter \\percentage} & \tabincell{l}{\% of uppercase letters in a post} & 1 \\
    \hline
    Special characters & \tabincell{l}{freq. of special characters} & 21 \\
    \hline
    Word shape & \tabincell{l}{freq. of all uppercase words, all \\lowercase words,
                                first character \\uppercase words,
                                camel case words} & 21 \\
    \hline
    \hline
    \tabincell{l}{Punctuation  freq.} & \tabincell{l}{freq. of punctuation, e.g., !,;?} & 10 \\
    \hline
    Function words & \tabincell{l}{freq. of function words} & 337 \\
    \hline
    POS tags & \tabincell{l}{freq. of POS tags, e.g., NP, JJ} & $<2300$ \\
    \hline
    \tabincell{l}{POS tag bigrams} & \tabincell{l}{freq. of POS tag bigrams} & $<2300^2$ \\
    \hline
    \hline
    \tabincell{l}{Misspelled words} & \tabincell{l}{freq. of misspellings} & 248 \\
    \hline
  \end{tabular}
  \end{table}
\end{center}

We extract various stylometric features from the WebMD and
HB datasets as shown in Table \ref{tab_feature}.
Generally, the features in Table \ref{tab_feature} can be classified
into three groups: \emph{lexical} features, \emph{syntactic} features,
and \emph{idiosyncratic} features.
The lexical features include length, word length, vocabulary richness,
letter frequency, digit frequency, uppercase letter percentage,
special characters, and word shape. They measure the writing style
of users with respect to characteristics of employed characters, words, and vocabularies.
The syntactic features include punctuation frequency, function words,
POS tags, and POS tag bigrams. They measure the writing style of users
with respect to the arrangement of words and phrases to create
well-formed sentences in posts.
For idiosyncratic features, we consider misspelled words,
which measure some peculiar writing style of users.

Since the number of POS tags and POS tag bigrams
could be variable, the number of total features is denoted by a variable $M$
for convenience.
According to the feature descriptions, all the features are real and positive valued.
Without loss of generality, we organize the features as a vector,
denoted by $\mathbf{F} = <F_1, F_2, \cdots, F_M>$.
Then, given a post, we extract its features with respect to $\mathbf{F}$
and obtain a feature vector consisting of 0 and positive real values,
where 0 implies that this post does not have the corresponding feature
while a positive real value implies that this post has the corresponding feature.

Note that, it is possible to extract more stylometric features
from the WebMD/HB dataset, e.g., content features \cite{abbchetis08}.
However, in this paper, we mainly focus on developing an effective online health data
DA framework. For the feature extraction part,
we mainly employ the existing techniques such as those in \cite{abbchetis08}-\cite{breafrtissec12},
and thus we do not consider this part as the technical contribution
of this paper. Certainly, understanding
which features are more effective in de-anonymizing online health data
is an interesting topic to study.
We take this as the future work of this paper.

\textbf{User-Data-Attribute Graph and Structural Features.}
Previously, we constructed a correlation graph $G$ for the users in
a health dataset. Now, we extend $G$ to a User-Data-Attribute (UDA) graph.
As the stylometric features demonstrate the writing characteristics of users,
logically, they can also be considered as the \emph{attributes} of users, which are similar to the
social attributes of users, e.g., career, gender, citizenship.
Therefore, at the user level, we define an \emph{attribute set/space}, denoted by $A$,
based on $\mathbf{F}$, i.e., $A = \{A_i | A_i = F_i, i = 1, 2, \cdots, M\}$.
Then, following this idea, for each feature $F_i \in \mathbf{F}$,
if a user $u$ has a post that has feature $F_i$ (i.e., the $F_i$ dimension is not 0
in the feature vector of that post), we say $u$ has attribute $A_i$, denoted by $u \sim A_i$.
Note that, each attribute is actually \emph{binary} to a user, i.e., a user either has an atrribute
$A_i$ or not, which is different from the feature, which could be
either a continuous or a discrete real value.
We define $A(u)$ as the set of all the attributes that user $u$ has,
i.e., $A(u) = \{A_i | A_i \in A, u\sim A_i\}$.
Since $u$ may have multiple posts that have feature $F_i$,
we assign a \emph{weight} to the relation $u \sim A_i$,
denoted by $l_u(A_i)$,
which is defined as the number of posts authored by $u$
that have the feature $F_i$.

Based on the attribute definition, we extend the correlation graph
to the UDA graph, denoted by $G = (V, E, W, A, O, L)$,
where $V$, $E$, and $W$ are the same as defined before,
$A$ is the attribute set, $O = \{u \sim A_i | u \in V, A_i \in A\}$
denotes the set of all the user-attribute relationships,
and $L = \{l_u(A_i) | u \sim A_i \in O\}$ denotes the set of
the user-attribute relationship weights.
Since the UDA graph
is an extension of the correlation graph, we use the same notation $G$ for these
two concepts.
In practice, one may consider more attributes of a user, e.g., their social attributes
(user's social information) and behavioral attributes (user's activity pattern), when
defining $A$.


From the definition of the UDA graph $G$, we can see that
it takes into account the data's correlation as well as the data's
linguistic features (by introducing the concept of attribute in
a different way compared to the traditional manner \cite{abbchetis08}-\cite{breafrtissec12}).
We will introduce how to use the UDA graph to conduct the user-level
DA and analyze the benefits in the following section.
Before that, we introduce more user-level features from the health data
leveraging the UDA graph.

The features extracted from the UDA graph are classified as
\emph{structural features}, which can be partitioned into
three categories: \emph{local correlation features},
\emph{global correlation features}, and \emph{attribute features}.
The local correlation features include user degree (i.e., $d_u$ for $u \in V$),
weighted degree (i.e., $wd_u$), and NCS vector (i.e., $\mathbf{D}_i$).
Basically, the local correlation features measure the \emph{direct interactivity}
of a user in a health forum.

Given $u \in V$ and a subset $S \subseteq V$,
the global correlation features of $u$ are defined as the
the distances and weighted distances from $u$ to the users in $S$,
denoted by vectors $\mathbf{H}_u(S) = <h_{uv} | v \in S>$
and $\mathbf{WH}_u(S) = <wh_{uv} | v \in S>$, respectively.
Basically, the global correlation features measure
the \emph{indirect interactivity} of a user in a health dataset.

Based on $A(u)$ of $u \in V$, we introduce a new notation
to take into account the weight of each attribute of $u$.
We define $WA(u) = \{(A_i, l_u(A_i)) | A_i \in A(u)\}$.
Then, the attribute features of $u \in V$ are defined
as $A(u)$ and $WA(u)$.
The attribute features measure the linguistic features
of users in the form of binary attributes and weighted binary attributes.

The defined structural features are helpful in conducting
user-level DA. We show this in detail in the
De-Health framework as well as in the experimental evaluations.

\section{De-anonymization} \label{da}

%

\subsection{Preliminary}

Before this work, the privacy vulnerability of online health data was unclear,
e.g., the health/medical data generated by users
of WebMD and HB, to the best of our knowledge.
In this section, we present a novel two-phase DA attack
to online health data. The considered anonymized data, denoted by $\Delta_1$, are the data
generated from current online health services, e.g., WebMD and HB.
There are multiple applications of those anonymized online health data:
($i$) as indicated in the privacy policies of WebMD and HB,
the health data of their users can be shared with researchers
for multiple research and analytics tasks \cite{webmd}\cite{hb}; ($ii$) again, according to their
privacy policies, the data could be shared with commercial partners (e.g., insurance companies
and pharmaceutical companies) for multiple business purposes \cite{webmd}\cite{hb};
and ($iii$) the data might be publicly released for multiple
government and societal applications \cite{usgov}\cite{cagov}.
Considering various applications of the online health data,
our question is: \emph{can those data be de-anonymized to
the users of online health services and can they be linked to the users' real identities}?
We answer the first part of this question in this
section by presenting De-Health and discuss the second part in Section \ref{linkage}.

To de-anonymize the anonymized data $\Delta_1$,
we assume that the adversary\footnote{Here, the adversaries are defined as the
ones who want to compromise the privacy of the users in the anonymized dataset.
During the data sharing and publishing process (for research, business, and
other purposes), every data recipient could be an adversary.
In our paper, we focus on studying the potential privacy vulnerability
of online health data.}
can collect some auxiliary data, denoted by $\Delta_2$, from the same or other online health
service. According to our knowledge, this is possible in practice:
from the adversary's perspective, for some online health services,
e.g., HB, it is not difficult to collect data from them using
some intelligent crawling techniques;
for some other online health services with strict policies,
e.g. PatientsLikeMe \cite{patientslikeme}, an adversary can also
collect their data by combining intelligent crawling techniques
and anonymous communication techniques (e.g., Tor).
In this paper, we assume both $\Delta_1$ and $\Delta_2$
are generated from online health services like WebMD and HB.

After obtaining the anonymized data $\Delta_1$ and the auxiliary data $\Delta_2$,
we extract the features of the data
and transform them into an anonymized graph and an auxiliary graph,
denoted by $G_1 = (V_1, E_1, W_1, A_1, O_1, L_1)$
and $G_2 = (V_1, E_1, W_1, A_1, O_1, L_1)$, respectively, using the techniques
discussed in Section \ref{datacollection}.
When it is necessary, we use the subscript `1' and `2'
to distinguish the anonymized data/graph and the auxiliary data/graph.
Now, the DA of $\Delta_1$ leveraging $\Delta_2$
can be approximately defined as: for an anonymized (unknown) user $u \in V_1$,
seeking an auxiliary (known) user $v \in V_2$,
such that $u$ can be identified to $v$ (i.e., they correspond
to the same real world person), denoted by $u \rightarrow v$.
However, in practice, it is unclear whether $\Delta_1$ and $\Delta_2$
are generated by the same group of users, i.e., it is unknown whether
$V_1  \stackrel{?}{=} V_2$. Therefore, we define \emph{closed-world DA}
and \emph{open-world DA}.
When the users that generate $\Delta_1$ are a subset of the users that generate $\Delta_2$,
i.e., $V_1 \subseteq V_2$, the DA problem is a \emph{closed-world DA}
problem. Then, a successful DA is defined as
$u \in V_1, v \in V_2$, $u \rightarrow v$ and $u$ and $v$ correspond
to the same user.
When $V_1 \neq V_2$, the DA problem is an \emph{open-world DA}
problem. Let $V_o = V_1 \cap V_2$, the overlapping users between
$V_1$ and $V_2$.
Then, a successful DA is defined as
$u \in V_1, v \in V_2$, $u \rightarrow v$, $u$ and $v$ are in $V_o$,
and $u$ and $v$ correspond
to the same user;
or $u \rightarrow \bot$, if $u \notin V_o$, where $\bot$ represents \emph{not-existence}.
For $u \in V_1$ and $v \in V_2$, if $u$ and $v$ correspond to
the same real world user, we call $v$ the \emph{true mapping}
of $u$ in $V_2$.
In this paper, the presented De-Health framework works for both
the closed-world and the open-world situations.

\subsection{De-Health}


\begin{algorithm} \label{a_de}
\SetKwInOut{Input}{input}
\SetKwInOut{Output}{output}
\BlankLine

construct $G_1$ and $G_2$ from $\Delta_1$ and $\Delta_2$, respectively\;
\For{{\color{blue}every $u \in V_1$}}
{
    \For{{\color{blue}every $v \in V_2$}}
    {
        {\color{blue}compute the \emph{structural similarity} between $u$ and $v$, denoted by $s_{uv}$\;}
    }
}
{\color{blue}compute the Top-$K$ candidate set for each user $u \in V_1$,
denoted by $C_u = \{v_i| v_i \in V_2, i = 1, 2, \cdots, K\}$,
based on the structural similarity scores\;
filter $C_u$ using a \emph{threshold vector}\;}
%
\For{{\color{red}$u \in V_1$}}
{
{\color{red}leveraging the stylometric and
structural features of the users in $C_u$,
build a classifier,
using benchmark machine learning techniques (e.g., SMO)\;}
{\color{red}using the classifier to de-anonymize $u$\;}
}
\caption{\textbf{De-Health}}
\end{algorithm}

\textbf{Overview.}
In this subsection, we present the De-Health framework.
We show the high level idea of De-Health in Algorithm \ref{a_de}
and give the details later.
At a high level,
De-Health conducts user DA in two phases:
\emph{Top-$K$ DA} (line 2-6) and
\emph{refined DA} (line 7-9).
In the Top-$K$ DA phase, we mainly focus on
de-anonymizing each anonymized user $u \in V_1$ to a Top-$K$ candidate set,
denoted by $C_u = \{v_i| v_i \in V_2, i = 1, 2, \cdots, K\}$,
that consists of the $K$ most structurally similar auxiliary users
with the anonymized user (line 2-5).
Then, we optimize the Top-$K$ candidate set using a threshold vector
by eliminating some less likely candidates (line 6).
In the \emph{refined DA} phase, an anonymized user
will be de-anonymized to some user in the candidate set
using a benchmark machine learning model trained leveraging
both stylometric and structural features.
Note that, we do not limit the DA scenario to closed-world
or open-world. De-Health is designed to take both scenarios into consideration.

\textbf{Top-$K$ DA.}
Now, we discuss how to implement Top-$K$ DA and optimization (filtering).

\emph{Structural Similarity.}
Before we compute the Top-$K$ candidate set for each anonymized user,
we compute the \emph{structural similarity} between each anonymized
user $u \in V_1$ and each auxiliary user $v \in V_2$,
denoted by $s_{uv}$,
from the graph perspective (line 2-3 in Algorithm \ref{a_de}).
In De-Health, $s_{uv}$ consists of three components:
\emph{degree similarity} $s^d_{uv}$, \emph{distance similarity} $s^s_{uv}$,
and \emph{attribute similarity} $s^a_{uv}$.
Specifically, $s^d_{uv}$ is defined as
\begin{align} \nonumber
s^d_{uv} = \frac{\min\{d_u, d_v\}}{\max\{d_u, d_v\}}
        + \frac{\min\{wd_u, wd_v\}}{\max\{wd_u, wd_v\}}
        + \cos(\mathbf{D}_u, \mathbf{D}_v),
\end{align}
where $\cos(\cdot, \cdot)$ is the \emph{cosine similarity}
between two vectors. Note that, it is possible that
$\mathbf{D}_u$ and $\mathbf{D}_v$ have different lengths.
In that case, we pad the short vector with zeros to ensure that both have the same length.
From the definition, $s^d_{uv}$ measures the degree similarity
of $u$ and $v$ in $G_1$ and $G_2$, i.e., their local direct interactivity
similarity in $\Delta_1$ and $\Delta_2$, respectively.

To define $s^s_{uv}$, we need to specify a set of \emph{landmark users}
from $G_1$ and $G_2$, respectively.
Usually, the landmark users are some \emph{pre-de-anonymized} users that serve
as \emph{seeds} for a DA \cite{narshmsp09}\cite{nilkapccs14}\cite{jiliccs14}.
There are many techniques to find landmark users,
e.g., clique-based technique \cite{narshmsp09}, community-based technique \cite{nilkapccs14},
and optimization-based technique \cite{jiliccs14}.
In De-Health, we do not require accurate landmark users.
In particular, we select $\hbar$ users with the largest degrees from $V_1$ and $V_2$
as the landmark users, denoted by $S_1$ and $S_2$, respectively.
We also sort the users in $S_1$ and $S_2$ in the degree decreasing order.
Then, we define $s^s_{uv}$ as
\begin{align} \nonumber
s^s_{uv} = \cos(\textbf{H}_u(S_1), \textbf{H}_v(S_2)) + \cos(\textbf{WH}_u(S_1), \textbf{WH}_v(S_2)).
\end{align}
Basically, $s^s_{uv}$ measures the relative global structural similarity,
i.e., indirect interactivity similarity, of $u$ and $v$.

For $u$ and $v$, we define $WA(u) \cap WA(v) = \{(A_i, l_{u\cap v}(A_i)) | A_i \in A(u) \cap A(v),
l_{u\cap v}(A_i) = \min\{l_u(A_i), l_v(A_i)\}\}$ and
$WA(u) \cup WA(v) = \{(A_i, l_{u\cup v}(A_i)) | A_i \in A(u) \cup A(v),
l_{u\cup v}(A_i) = \max\{l_u(A_i), l_v(A_i)\}\}$.
Further, let $|\cdot|$ be the cardinality of a set
and for the weighted set, we define
$|WA(u) \cap WA(v)| = \sum\limits_{A_i \in A(u) \cap A(v)} l_{u\cap v}(A_i)$
and $|WA(u) \cup WA(v)| = \sum\limits_{A_i \in A(u) \cup A(v)} l_{u\cup v}(A_i)$.
Then, $s^a_{uv}$ is defined as
\begin{align} \nonumber
s^a_{uv} = \frac{|A(u) \cap A(v)|}{|A(u) \cup A(v)|} + \frac{|WA(u) \cap WA(v)|}{|WA(u) \cup WA(v)|},
\end{align}
which measures the attribute similarity (i.e., linguistic similarity)
between $u$ and $v$.

After specifying $s^d_{uv}$, $s^s_{uv}$, and $s^a_{uv}$, the structural similarity
between $u$ and $v$ is defined as
\begin{align} \nonumber
s_{uv} = c_1 \cdot s^d_{uv} + c_2 \cdot s^s_{uv} + c_3 \cdot s^a_{uv},
\end{align}
where $c_1, c_2$ and $c_3$ are some positive constant values
adjusting the weights of each similarity component.

\emph{Top-$K$ Candidate Set.}
After obtaining the structural similarity scores, we compute the
Top-$K$ candidate set $C_u$ for each $u \in V_1$
(line 5 in Algorithm \ref{a_de})\footnote{Here,
we assume that $K$ is far less than the number of auxiliary users.
Otherwise, it is meaningless to seek Top-$K$ candidate sets.}.
Here, we propose two approaches: \emph{direct selection}
and \emph{graph matching based selection}.
In \emph{direct selection}, we directly select $K$ auxiliary users
from $V_2$ that have the Top-$K$ similarity scores with $u$.
In \emph{graph matching based selection}: Step 1: we first construct a \emph{weighted completely
connected bipartite graph} $G(V_1, V_2)$ (anonymized users on one side while
auxiliary users on the other side), where the weight on each edge
is the structural similarity score between the two corresponding users;
Step 2: we find a \emph{maximum weighted bipartite graph matching} on $G(V_1, V_2)$,
denoted by $\{(u_i, v_i) | u_i \in V_1, v_i, i = 1, 2, \cdots, |V_1|\}$;
Step 3: for each $(u_i, v_i)$ in the matching, we add $v_i$ to the Top-$K$
candidate set of $u_i$ and remove the edge between $u_i$ and $v_i$
in the bipartite graph $G(V_1, V_2)$;
Step 4: repeat Steps 2 and 3 until we find a Top-$K$ candidate set for each user
in $V_1$.

\begin{algorithm} \label{a_filter}
\SetKwInOut{Input}{input}
\SetKwInOut{Output}{output}
\BlankLine
$s_{u} \leftarrow \max\{s_{uv} | u \in V_1, v \in V_2\}$\;
$s_{l} \leftarrow \min\{s_{uv} | u \in V_1, v \in V_2\} + \epsilon$\;
construct a \emph{threshold vector} $\mathbf{T} = <T_i>$,
where for $i = 0, 1, \cdots, l-1>$, $T_i = s_{u} - \frac{i}{l - 1} \cdot (s_u - s_l)$\;
\For{every $u \in V_1$}
{
    \For{$i = 0; i \leq l - 1; i++$}
    {
        $C_u' \leftarrow C_u$\;
        \For{$v \in C_u'$}
        {
            \If{$s_{uv} < T_i$}
            {
                $C_u' = C_u' \setminus \{v\}$\;
            }
        }
        \If{$C_u' \neq \emptyset$}
        {
            $C_u \leftarrow C_u'$,
            \textbf{break}\;
        }
    }
    \If{$C_u' = \emptyset$}
    {
        $u \rightarrow \bot$,
        $V_1 \leftarrow V_1 \setminus \{u\}$\;
    }
}
\caption{\textbf{Filtering}}
\end{algorithm}

\emph{Optimization/Filtering.} After determining the Top-$K$ candidate set
for each $u \in V_1$, we further optimize $C_u$ using the
\emph{filtering procedure} shown in Algorithm \ref{a_filter}
(to finish line 6 in Algorithm \ref{a_de}),
where $\epsilon \in [0, s_u - \min\{s_{uv} | u \in V_1, v \in V_2\}]$
is a positive constant value, $l$ is the length of the threshold vector $\mathbf{T}$
(defined later), and $C'_u$ is a temporary candidate set.
The main idea of the filtering process is to pre-eliminate some
less likely candidates in terms of structural similarity
using a threshold vector.
Below, we explain Algorithm \ref{a_filter} in detail.
First, the threshold interval $[s_l, s_u]$ is specified based on $\epsilon$, and the maximum
and minimum similarity scores between the users in $V_1$ and $V_2$ (line 1-2).
Then, the threshold interval is partitioned into $l$ segments
with the threshold value $T_i = s_{u} - \frac{i}{l - 1} \cdot (s_u - s_l)$
($i = 0, 1, \cdots, l-1$). We organize the threshold values
as a \emph{threshold vector} $\mathbf{T} = <T_i>$ (line 3).
Third, we use $\mathbf{T}$ to filter each candidate set $C_u$
starting from large thresholds to small thresholds (line 5-13).
If some candidate users pass the filtering at some threshold level,
we then break the filtering process and take those candidate users
as the final $C_u$ (line 7-10).
If no candidate users are left even after being filtered by $T_{l-1}$
(the smallest threshold), we conclude that $u$ does not appear
in the auxiliary data (i.e., $u \rightarrow \bot$) and remove
$u$ from $V_1$ for further consideration (line 12-13).

Note that, the filtering process is mainly used for reducing the
size of the candidate set for each anonymized user,
and thus to help obtain a better refined DA result
and accelerate the DA process in the following stage.
In practice, there is no guarantee for the filtering to improve the
DA performance.
Therefore, we set the filtering process
as an \emph{optional choice} for De-Health.

\textbf{Refined DA.}
In the first phase of De-Health, we seek a Top-$K$ candidate set
for each anonymized user. In the second phase (line 7-9 of Algorithm \ref{a_de}),
De-Health conducts
refined DA for each $u \in V_1$ and either de-anonymizes
$u$ to some auxiliary user in $C_u$ or concludes that $u \rightarrow \bot$,
i.e., $u$ does not appear in the auxiliary data.
To fulfill this task, the high level idea is:
leveraging the stylometric and correlation features of the users
in $C_u$, train a classifier employing benchmark machine learning techniques,
e.g., Support Vector Machine (SVM), Nearest Neighbor (NN),
Regularized Least Squares Classification (RLSC),
which is similar to that in existing stylometric approaches
\cite{abbchetis08}-\cite{calharusenix15}\footnote{In \cite{abbchetis08}-\cite{calharusenix15},
multiple benchmark machine learning based stylometric approaches are proposed
to address the post/passage-level author attribution.
Although we focus on user-level DA, those approaches could
be extended to our refined DA phase.}.
Therefore, we do not go to further details to explain
existing benchmark machine learning techniques.

Nevertheless, there is still an open problem here: by default,
existing benchmark machine learning techniques are satisfiable
at addressing the closed-world DA problem
(e.g., \cite{abbchetis08}\cite{narpassp12}). However, their performance
is far from expected in open-world DA \cite{stoovewg14}.
To address this issue, we present two schemes:
\emph{false addition} and \emph{mean-verification},
which are motivated by the open-world author attribution techniques
proposed by Stolerman et al. in \cite{stoovewg14}.

In the \emph{false addition} scheme, when de-anonymizing $u \in V_1$,
we randomly select $K'$ users from $V_2 \setminus C_u$ (e.g., $K' = |C_u|$),
and add these $K'$ users to $C_u$ as \emph{false users}.
Then, if $u$ is de-anonymized to a false user in $C_u$,
we conclude that $u \rightarrow \bot$, i.e., $u$ does not appear in
the auxiliary data. Otherwise, $u$ is de-anonymized to a non-false user.

In the \emph{mean-verification} scheme, we first use the trained classifier to de-anonymize
$u$ to some user, say $v$, in $C_u$ by assuming it is a closed-world DA problem.
Later, we verify this DA:
let $\lambda_u = (\sum\limits_{w \in C_u} s_{uw})/|C_u|$ be the \emph{mean similarity}
between $u$ and its candidate users;
then, if $s_{uv} \geq (1 + r) \cdot \lambda_u$, where $r \geq 0$ is some predefined
constant value,
the DA $u \rightarrow v$
is accepted; otherwise, it is rejected, i.e., $u \rightarrow \bot$.
Note that, the verification process can also be implemented using other
techniques, e.g., distractorless verification \cite{noeryalrec12}, Sigma verification \cite{stoovewg14}.

\textbf{Remark.}
%
To the best of knowledge, De-Health is the \emph{first} user-level DA attack
on online health data. In De-Health, we propose a novel approach to
construct a \emph{UDA graph} based on the health data by systematically
characterizing the interactivity correlations among different users as well as
the writing characteristics of users. The UDA graph further enables us to
develop effective graph-based DA techniques, which can be easily
scaled to large-scale data. Moreover, the UDA graph enables us to extract
various structural features, which can be used to feed benchmark machine learning
techniques together with traditional stylometric features
and train more effective classifiers.

The Top-$K$ DA phase of De-Health can improve
the DA performance from multiple perspectives.
On one hand, it significantly reduces the possible mapping space
for each anonymized user (from $V_2$ to $C_u$),
and thus a more accurate classifier
can be trained to de-anonymize an anonymized user, followed
by the improved DA performance.
From the description of De-Health (Algorithm \ref{a_de}),
it seems that the Top-$K$ DA might
degrade its DA performance if many
true mappings
of the anonymized users cannot be included into their Top-$K$ candidate sets.
However, we seek the candidate set for each anonymized user $u$ based on structure similarities
between $u$ and the users in $V_2$, and the auxiliary users
that have high structural similarities with $u$ are preferred
to be selected as candidates,
e.g., in the direct selection approach.
According to our theoretical analysis in the following section,
\emph{this candidate selection approach will not degrade the DA
performance} in practice.
Furthermore, as shown in our experiments (Section \ref{experiment}),
most anonymized users' true mappings can be selected into their candidate sets
when a proper $K$ is chosen.
On the other hand, since the possible mapping space is significantly
reduced by the Top-$K$ DA, the computational cost
for both constructing the machine learning based classifiers
and performing refined DA can be reduced.

Most real world DA tasks are open-world problems.
By introducing the false addition scheme and the mean-verification scheme,
De-Health can address both closed-world and open-world DA
issues.

\section{Theoretical Analysis} \label{theory}

In this section, we present a general theoretical analysis framework
for the soundness and effectiveness of online health data DA,
which can also serve as the theoretical foundation of De-Health.

\subsection{Preliminary}

For the convenience of analysis, we introduce some formal notations.
We assume $\Delta_1$ is an anonymized online health dataset generated
by $n_1$ users denoted by set $V_1$
and $\Delta_2$ is an auxiliary dataset generated by
$n_2$ users denoted by set $V_2$. Note that, it is possible that $n_1 \neq n_2$.
However, since we employ $\Delta_2$ to de-anonymize $\Delta_1$,
we assume there are some overlap between users in $\Delta_1$ and $\Delta_2$.
Otherwise, this DA is meaningless.
Let $u \in \Delta_1$ be an overlapping anonymized user and $u' \in \Delta_2$ be the
true mapping of $u$ ($u$ and $u'$ correspond to the same real world people).

To de-anonymize $\Delta_1$ leveraging $\Delta_2$, many features of the data will
be extracted to develop a DA algorithm/model.
The \emph{feature} here is a general concept,
which could include stylometric features, structural features, social features,
and other possible features in our theoretical analysis.
Thus, we define a \emph{general feature space} $\mathcal{F}$
to characterize all the possible features, attributes, and other measurements
of users. Then, given a user $u$, we denote its features by a feature vector
$\mathcal{F}_u$. Based on the features of $\Delta_1$
and $\Delta_2$, we construct
a DA model/algorithm, denoted by $\mathcal{M}$.
For instance, De-Health can be considered as one implementation of $\mathcal{M}$:
it de-anonymizes $\Delta_1$ by employing structural similarity (derived from
the defined structural features), stylometric features, and correlation features.
Ideally, $\mathcal{M}$ works in the following manner:
if $u \in V_1$ is an overlapping user, we have $\mathcal{M}(u, V_2): u \rightarrow u'$,
i.e., $\mathcal{M}$ successfully de-anonymizes $u$ to $u'$;
otherwise, we have $\mathcal{M}(u, V_2): u \rightarrow \bot$.

To design $\mathcal{M}$, we introduce a general function $f(\cdot, \cdot)$,
which is defined on the features of two users (e.g., $f(\mathcal{F}_u, \mathcal{F}_v)$
for $u \in V_1$ and $v \in V_2$) and measures
the \emph{distance} of the two users in terms of their features.
Note that, the \emph{distance} concept here is very general.
It could be defined in terms of different metrics,
e.g., the distribution similarity or the Euclidian
distance between the feature vectors of two users,
depending on a particular DA algorithm.
For instance, when $f(\cdot, \cdot)$ is defined using the feature
distribution similarity, it can be defined as a decreasing function
with respect to the distribution similarity, i.e.,
a higher similarity implies a smaller distance ($f(\cdot, \cdot)$ value).
Using $f(\cdot, \cdot)$,
mathematically, $\mathcal{M}$ can also be constructed as a \emph{function}.
For instance, we can define $\mathcal{M}$ as
$\mathcal{M}(u \in V_1, v \in V_2) = p = f(\mathcal{F}_u, \mathcal{F}_v) /
\sum\limits_{x \in V_2} f(\mathcal{F}_u, \mathcal{F}_x)$\footnote{Here,
an implicated assumption is that $f(\cdot, \cdot) \geq 0$ and
$\exists x \in V_2$ such that $f(\mathcal{F}_u, \mathcal{F}_x) \neq 0$.
Since $f(\cdot, \cdot)$ is a distance function, this assumption is intuitively
reasonable in practice.}, where $p \in [0, 1]$
indicates the \emph{probability} that $u$ is de-anonymized to $v$
by $\mathcal{M}$.

Let $\lambda$ be the \emph{mean value} of correct DAs
under $f(\cdot, \cdot)$, i.e., $\lambda = \mathbf{E}[f(u, u')]$
\footnote{$\mathbf{E}[\cdot]$ denotes the \emph{mean/expectation} value in this paper.},
and $\overline{\lambda}$ be \emph{mean value} of incorrect DAs
under $f(\cdot, \cdot)$, i.e., $\overline{\lambda} = \mathbf{E}[f(u, v)]$
where $v \in V_2$ and $v \neq u'$.
Furthermore, assume that $f(u, u') \in [\theta_l, \theta_u]$
and $f(u, v) \in [\overline{\theta_l}, \overline{\theta_u}]$ ($v \neq u'$),
i.e., the \emph{ranges} of correct and incorrect DAs
under $f$ are $[\theta_l, \theta_u]$ and $[\overline{\theta_l}, \overline{\theta_u}]$,
respectively. Let $\theta = \theta_u - \theta_l$, $\overline{\theta} = \overline{\theta_u}
- \overline{\theta_l}$, and $\delta = \max\{\theta, \overline{\theta}\}$.
Below, we analyze the
\emph{re-identifiability}, defined as the probability of being successfully de-anonymized,
of $\Delta_1$, and further specify the design of a corresponding $\mathcal{M}$
to achieve that re-identifiability.

Due to the space limitations, we place all the proofs
for theorems and corollaries in Appendix \ref{proof}.

\subsection{Re-Identifiability Analysis}

We start the re-identifiablity analysis from the simple case that:
given $u \in V_1$ be an overlapping user, $u', v \in V_2$, and $v \neq u'$,
deriving the re-identifiability of $u$ with respect to $\{u', v\}$.
Let $\Pr(u \stackrel{\{u', v\}}{\rightarrow} u')$
be the probability of $\exists \mathcal{M}$ such that $\mathcal{M}$
can successfully de-anonymize $u$ to $u'$ from $\{u', v\}$.
Then, we have the following theorem on quantifying $\Pr(u \stackrel{\{u', v\}}{\rightarrow} u')$.
We also specify the design of $\mathcal{M}$ in the proof.

\begin{theorem} \label{t1}
When $\lambda \neq \overline{\lambda}$,
$\Pr(u \stackrel{\{u', v\}}{\rightarrow} u') \geq 1 - 2\exp(-\frac{(\lambda - \overline{\lambda})^2}
{4 \delta^2})$.
\end{theorem}

In Theorem \ref{t1}, we derived the probability of successfully
de-anonymizing $u$ from $\{u', v\}$ and gave the design of $\mathcal{M}$.
Based on Theorem \ref{t1},
we can obtain a stronger conclusion as shown in Corollary \ref{c1}
using stochastic theory,
which states the asymptotical property of the DA.
In the corollary, $n$ is a positive integer and the same $\mathcal{M}$
as specified in Theorem \ref{t1} is employed.

\begin{corollary} \label{c1}
When $\lambda \neq \overline{\lambda}$ and $|\lambda - \overline{\lambda}|/2\theta
\geq \sqrt{2\ln n + \ln 2}$, $\Pr(u \stackrel{\{u', v\}}{\rightarrow} u')
\stackrel{n \rightarrow \infty}{\rightarrow}  1$, i.e., it is asymptotically almost surely (a.a.s.)
that $u$ can be successfully de-anonymized from $\{u', v\}$ \footnote{Asymptotically almost surely (a.a.s.)
implies that as $n \rightarrow \infty$, an event happens with probability goes to 1.}.
\end{corollary}


In Theorem \ref{t1}, we studied the re-identifiability of
de-anonymizing $u$ from $\{u', v\}$.
In practice, as shown in De-Health, we need to de-anonymize
$u$ from $V_2$, the set of all the auxiliary users.
We give the re-identifiability of $u \in V_1$ in this general
case in Corollary \ref{c2}.
Now, suppose that $u' \in V_2$, i.e., $u$ is an overlapping user of
$V_1$ and $V_2$.
We define $\Pr(u \stackrel{V_2}{\rightarrow} u')$
as the probability of $\exists \mathcal{M}$ such that $\mathcal{M}$
can successfully de-anonymize $u$ to $u'$ from $V_2$
($\mathcal{M}$ is specified in the proof).

\begin{corollary} \label{c2}
When $\lambda \neq \overline{\lambda}$ and $|\lambda - \overline{\lambda}|/2\theta
\geq \sqrt{2\ln n + \ln 2n_2}$, $\Pr(u \stackrel{V_2}{\rightarrow} u')
\stackrel{n \rightarrow \infty}{\rightarrow}  1$, i.e., it is a.a.s.
that $u$ can be successfully de-anonymized from $V_2$.
\end{corollary}

Corollary \ref{c2} is an even stronger conclusion than that
in Corollary \ref{c1}. It specifies the conditions to successfully
de-anonymize an anonymized user in general.

Now, we study the re-identifiability of any subset of $V_1$,
i.e., a subset of anonymized users.
Let $\alpha \in [0, 1]$ be some constant value
and $\alpha n_1$ be an integer.
For $\forall V_\alpha \subseteq V_1$, it is an \emph{$\alpha$-subset}
if $|V_\alpha| = \alpha n_1$
and $\forall u \in V_\alpha$, $u$ has a true mapping $u'$ in $V_2$.
Then, we define that $\Delta_1$ is \emph{$\alpha$-re-identifiable}
if there exists an $\alpha$-subset $V_\alpha$ of $V_1$ such that
$\forall u \in V_\alpha$, $\Pr(u \stackrel{V_2}{\rightarrow} u')
\stackrel{n \rightarrow \infty}{\rightarrow}  1$.
We give the probability that $\Delta_1$ is $\alpha$-re-identifiable
in the following theorem.

\begin{theorem} \label{t2}
Suppose that $\Delta_1$ has an $\alpha$-subset $V_\alpha$.
Then, when $\lambda \neq \overline{\lambda}$,
$\Pr(\Delta_1$ is $\alpha$-re-identifiable$) \geq 1 - \exp(\ln 2 \alpha n_1 n_2
- \frac{(\lambda - \overline{\lambda})^2}{4 \theta^2})$.
\end{theorem}

In Theorem \ref{t2}, we derived the probability that $\Delta_1$ is
$\alpha$-re-identifiable. Similar to that in Corollary \ref{c2},
we now derive the conditions to have $\Delta_1$ stochastically
$\alpha$-re-identifiable. We show the result in Corollary \ref{c3}.
In the proof, we use the same $\mathcal{M}$ design as that in
Theorem \ref{t2}.

\begin{corollary} \label{c3}
Suppose that $\Delta_1$ has an $\alpha$-subset $V_\alpha$.
Then, when $\lambda \neq \overline{\lambda}$ and
$|\lambda - \overline{\lambda}|/2\theta \geq \sqrt{2\ln n + \ln 2 \alpha n_1 n_2}$,
$\Pr(\Delta_1$ is $\alpha$-re-identifiable$)
\stackrel{n \rightarrow \infty}{\rightarrow}  1$, i.e., it is a.a.s.
that $\Delta_1$ is $\alpha$-re-identifiable.
\end{corollary}


In Theorem \ref{t2} and Corollary \ref{c3}, we derived the probability
to have $\Delta_1$ $\alpha$-re-identifiable as well as the
conditions for $\Delta_1$ to be $\alpha$-re-identifiable.
Since $\alpha \in [0, 1]$, those results provide the theoretical analysis
for general online health data DA.

\subsection{Top-$K$ Re-Identifiability Analysis}

In the previous subsection, we analyzed the probability of de-anonymizing
one user or a group of users. We also derived the conditions
to have one or a group of users to be a.a.s. re-identifiable.
In the DA research,
in addition to accurate DA,
we may also have interest in understanding the probability/conditions
for conducting Top-$K$ DA.
Thus, we give the Top-$K$ re-identifiability analysis
in this subsection.

Formally, for $u \in V_1$, suppose it has a true mapping $u' \in V_2$.
Then, a correct Top-$K$ DA of $u$ is to seek a candidate set $C_u$
for $u$ such that $C_u \subseteq V_2$, $|C_u| = K$ (in this paper,
it is also acceptable if $|C_u| < K$), and $u' \in C_u$.
Let $\Pr(u \rightarrow C_u)$ be the probability of that $\exists \mathcal{M}$
and $\mathcal{M}$ can find a correct Top-$K$ candidate set $C_u$ for $u$.
We show the Top-$K$ re-identifiability of one user
and the conditions to have it asymptotically Top-$K$ re-identifiable
in the following theorem.

\begin{theorem} \label{t3}
When $\lambda \neq \overline{\lambda}$, ($i$)
$\Pr(u \rightarrow C_u) \geq 1 -
\exp(\ln 2 (n_2 - K) -\frac{(\lambda - \overline{\lambda})^2} {4 \delta^2})$;
($ii$) if $|\lambda - \overline{\lambda}|/2\theta \geq \sqrt{\ln 2(n_2 - K) + 2 \ln n}$,
$\Pr(u \rightarrow C_u)
\stackrel{n \rightarrow \infty}{\rightarrow}  1$, i.e., it is a.a.s.
that $u$ is Top-$K$ re-identifiable.
\end{theorem}

In Theorem \ref{t3}, we derived the Top-$K$ re-identifiability of a user
and the conditions to asymptotically de-anonymize the user.
Now, we extend our analysis to a general scenario of
Top-$K$ DA of a set of anonymized users.
Let $V_\alpha$ be an $\alpha$-subset of $\Delta_1$.
Then, we define $\Delta_1$ to be \emph{Top-$K$ $\alpha$-re-identifiable}
if $\forall u \in V_\alpha$, $\exists \mathcal{M}$ and $\mathcal{M}$
can find a correct Top-$K$ candidate set $C_u$ for $u$.
Let $\Pr(V_\alpha: u \rightarrow C_u)$ be the probability that
$\Delta_1$ is Top-$K$ $\alpha$-re-identifiable.
Then, we show the Top-$K$ re-identifiability of $V_\alpha$
and the conditions to have it asymptotically Top-$K$ re-identifiable
in the following theorem.

\begin{theorem} \label{t4}
When $\lambda \neq \overline{\lambda}$, ($i$)
$\Pr(V_\alpha: u \rightarrow C_u) \geq 1 -
\exp(\ln 2 \alpha n_1 (n_2 - K) -\frac{(\lambda - \overline{\lambda})^2} {4 \delta^2})$;
($ii$) if $|\lambda - \overline{\lambda}|/2\theta \geq \sqrt{\ln 2\alpha n_1 (n_2 - K) + 2 \ln n}$,
$\Pr(V_\alpha: u \rightarrow C_u)
\stackrel{n \rightarrow \infty}{\rightarrow}  1$, i.e., it is a.a.s.
that $\Delta_1$ is Top-$K$ $\alpha$-re-identifiable.
\end{theorem}

\section{Experiments} \label{experiment}

In this section, we experimentally evaluate De-Health
leveraging the two collected online health datasets: WebMD and HB.
First, we evaluate De-Health's performance in the
\emph{closed-world DA} setting, i.e., for each anonymized user,
its true mapping is in the auxiliary data (training data).
Then, we extend our evaluation to the more practical
\emph{open-world DA} setting: for each anonymized user,
its true mapping may or may not appear in the auxiliary data.

\subsection{Closed-world DA}


\subsubsection{Top-$K$ DA.}

First, we evaluate the Top-$K$ DA performance of De-Health.
In the Top-$K$ DA phase, we seek a candidate set $C_u \subseteq V_2$
for each anonymized user $u$. We define that the Top-$K$ DA of
$u$ is \emph{successful/correct} if $u$'s true mapping is included in the $C_u$
returned by De-Health.
Note that, the Top-$K$ DA is crucial to the success and overall performance
of De-Health: given a relatively large auxiliary dataset
and a small $K$, if there is a high success rate
in this phase, the candidate space of finding the
true mapping of an anonymized user can be significantly reduced
(e.g., from millions or hundreds of thousands of candidates to several hundreds of candidates).
Then, many benchmark machine learning techniques can
be employed to conduct
the second phase refined (precise) DA, since
as shown in \cite{abbchetis08}-\cite{calharusenix15},
benchmark machine learning techniques can achieve much better performance
on a relatively small training dataset
than on a large training dataset\footnote{In the closed-world
author attribution setting,
state-of-the-art machine learning based stylometric approaches
can achieve $\sim 80\%$ accuracy on 100-level of users \cite{abbchetis08},
$\sim 30\%$ accuracy on 10K-level of users \cite{kopschlre11},
and $\sim 20\%$ accuracy on 100K-level of users \cite{narpassp12}.}.

\textbf{Methodology and Setting.}
We partition each user's data (posts) in WebMD and HB into two
parts: \emph{auxiliary data} denoted by $\Delta_2$ and
\emph{anonymized data} denoted by $\Delta_1$.
Specifically, we consider three scenarios:
randomly taking $50\%$, $70\%$, and $90\%$ of each user's data
as auxiliary data and the rest as anonymized data
(by replacing each username with some random ID), respectively.
Then, we run De-Health to identify a Top-$K$ candidate
set for each user in $\Delta_1$ and examine the CDF
of the successful Top-$K$ DA with respect to
the increase of $K$.
For the parameters in De-Health, the default settings
are: $c_1 = 0.05$, $c_2 = 0.05$, and $c_3 = 0.9$.
We assign low weights to degree and distance similarities
when computing the structural similarity.
This is because, as shown in Section \ref{datacollection},
even in the UDA graph constructed based on the whole WebMD/HB dataset,
($i$) the degree of most of the users is low; and
($ii$) the size of most identified communities is small
and the UDA graph is disconnected (consisting of tens of disconnected components).
After partitioning the original dataset into auxiliary and anonymized data,
the degree of most users gets lower and the connectivity of the UDA graph
decreases further, especially in the scenario of $10\%$-anonymized data
(the anonymized UDA graph consists of hundreds of disconnected components
in our experiments).
Thus, intuitively, the degree and distance (vector) do not provide
much useful information in distinguishing different users for the
two leveraged datasets here, and we
assign low weights to degree and distance similarities.
Furthermore, we set the number of landmark users as $\hbar = 50$
(the Top-50 users with respect to degree).
For the structural similarity based Top-$K$ candidate selection,
we employ the \emph{direct selection approach}.
Since we conduct closed-world
evaluation in this subsection, the filtering process is omitted.
All the experiments are run 10 times. The results are the
average of those 10 runs.


\begin{figure}
\centering
\includegraphics[width=3.5in]{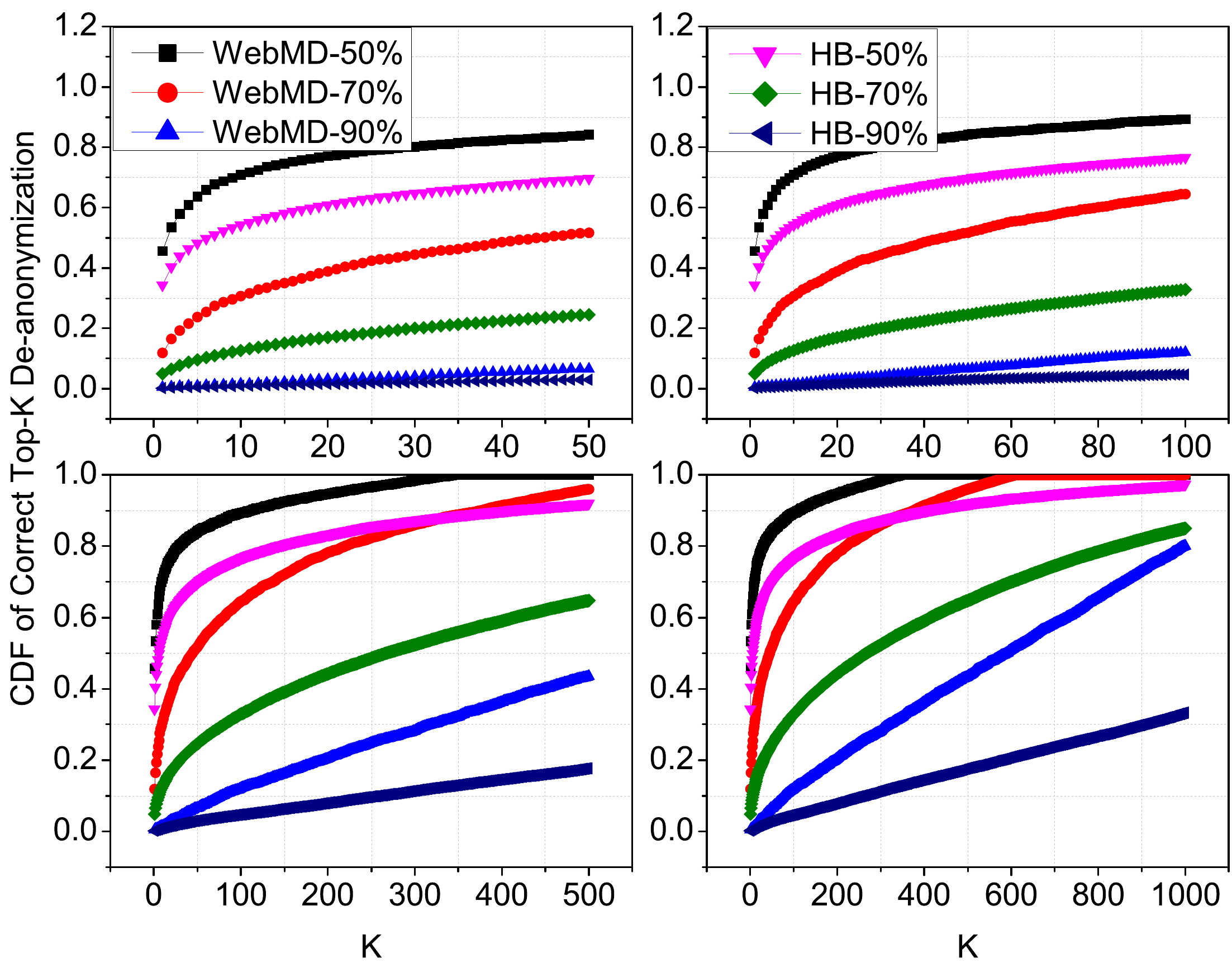}
\caption{CDF of correct Top-$K$ DA.} \label{f_topkc}
\end{figure}

\textbf{Results.}
We show the CDF of successful Top-$K$ DA with respect to
different $K$ ranges ($K \in [1, 50]$, $K \in [1, 100]$, $K \in [1, 500]$,
and $K \in [1, 1000]$) in Fig.\ref{f_topkc}. We have the following
observations.

First, with the increasing of $K$, the CDF of successful
Top-$K$ DA increases.
The reason is evident. When $K$ increases,
the probability of including the true mapping of an anonymized user
to its Top-$K$ candidate set also increases.

Second, when comparing the Top-$K$ DA performance of De-Health
on WebMD and HB, De-Health has a better performance on WebMD than that on HB.
For instance, when de-anonymizing the two datasets in the $70\%$-auxiliary data scenario,
De-Health finds the correct Top-500 candidate sets for $96\%$ WebMD users
while finds  the correct Top-500 candidate sets for $64.7\%$ HB users.
This is due to the fact that the HB dataset (388,398 users) has many more users
than the WebMD  dataset (89,393 users), and thus with a higher probability,
the correct Top-$K$ candidate set can be found for a WebMD user
under the same experimental setting.

Third, the size of the available dataset (either the auxiliary data
or the anonymized data) is important to constructing the UDA graph
and thus has an explicit impact on the Top-$K$ DA performance.
For instance, when de-anonymizing WebMD, De-Health can find
the correct Top-500 candidate sets for $100\%$ anonymized users
in the $50\%$-auxiliary data scenario
while can find  the correct Top-500 candidate sets for $43.5\%$ anonymized users
in the $90\%$-auxiliary data scenario.
This is because in the $90\%$-auxiliary data scenario, only $10\%$ of the original
dataset severs as the anonymized data. Then, only a very sparse anonymized UDA graph
that consists of hundreds of disconnected components can be constructed.
Thus, the Top-$K$ DA performance has been clearly degraded.

Overall, De-Health is powerful in conducting Top-$K$ DA
on large-scale datasets (especially,
when sufficient data appear in the auxiliary/anonymized data).
By seeking each anonymized user  Top-$K$ candidate set,
it decreases the DA space for a user from
100K-level to 100-level with high accuracy.
This is further very meaningful for the following up
refined DA, which enables the development of
an effective machine learning based classifier.

\subsubsection{Refined DA}
We have demonstrated the effectiveness of the Top-$K$ DA
of De-Health on large-scale datasets.
Now, we evaluate the refined DA phase of De-Health.
As we indicated in Section \ref{da}, the refined DA
can be implemented by training a classifier employing existing benchmark machine
learning techniques similar to those in \cite{abbchetis08}-\cite{calharusenix15}.
In addition, more than $96.6\%$ (resp., $98.2\%$) WebMD users
and more than $92.2\%$ (resp., $95.6\%$) HB users have less than 20 (resp., 40) posts,
and the average length of those posts is short
(the average lengths for WebMD posts and HB posts
are 127.59 words and 147.24 words, respectively).
Therefore, to enable the application of machine learning techniques
to train a meaningful classifier\footnote{As indicated in
\cite{abbchetis08}\cite{narpassp12}\cite{afrbresp12}\cite{calharusenix15},
when applying machine learning based stylometric approaches for author attribution,
there is a minimum requirement on the number of training words,
e.g., 4500 words and 7500 words, for obtaining a meaningful classifier.},
we conduct this group of evaluation on small-scale datasets extracted from the
WebMD dataset,
which is actually sufficient to show the performance of De-Health.

\textbf{Methodology and Settings.}
We construct the auxiliary (training) and anonymized (testing) data
for two evaluation settings. In the first setting, we randomly select
50 users each with 20 posts. Then, for the posts of each user, we take
10 for training (auxiliary data) and the other 10 (anonymized) for testing.
In the second setting, we randomly select 50 users each with 40 posts.
Then, we take 20 posts from each user for training (auxiliary data)
and take the remaining data for testing (anonymized). For each setting,
we conduct 10 groups of evaluations. The reported results are the average
of those 10 evaluations.

For the parameters in De-Health, the default settings are:
$c_1 = 0.05$, $c_2 = 0.05$, $c_3 = 0.9$ (the reason is the same as before),
$\hbar = 5$, $\epsilon = 0.01$, and $l = 10$;
the employed Top-$K$ candidate set selection approach
is \emph{direct selection}.
In the refined DA phase, the employed machine learning techniques
for training the classifier are the $k$-Nearest Neighbors (KNN) algorithm
\cite{narpassp12}
and the Sequential Minimal Optimization (SMO) Support Vector Machine \cite{stoovewg14}.
Note that, our settings and evaluations can be extended to other machine
learning techniques directly. The features used to train the classifier
are the stylometric features and structural features extracted
from the auxiliary data (as defined in Section \ref{datacollection}).

We also compare De-Health with a DA method that is similar
to traditional stylometric approaches \cite{abbchetis08}-\cite{breafrtissec12}:
leveraging the same feature set as in De-Health, training a classifier using KNN and SMO
without of our Top-$K$ DA phase, and employing the classifier
for DA. We denote this comparison method as \emph{Stylometry}
(although we included correlation features in addition to stylometric features).
Actually, Stylometry is equivalent to the second phase (refined DA)
of De-Health.


\begin{figure}
\centering
\includegraphics[width=2in]{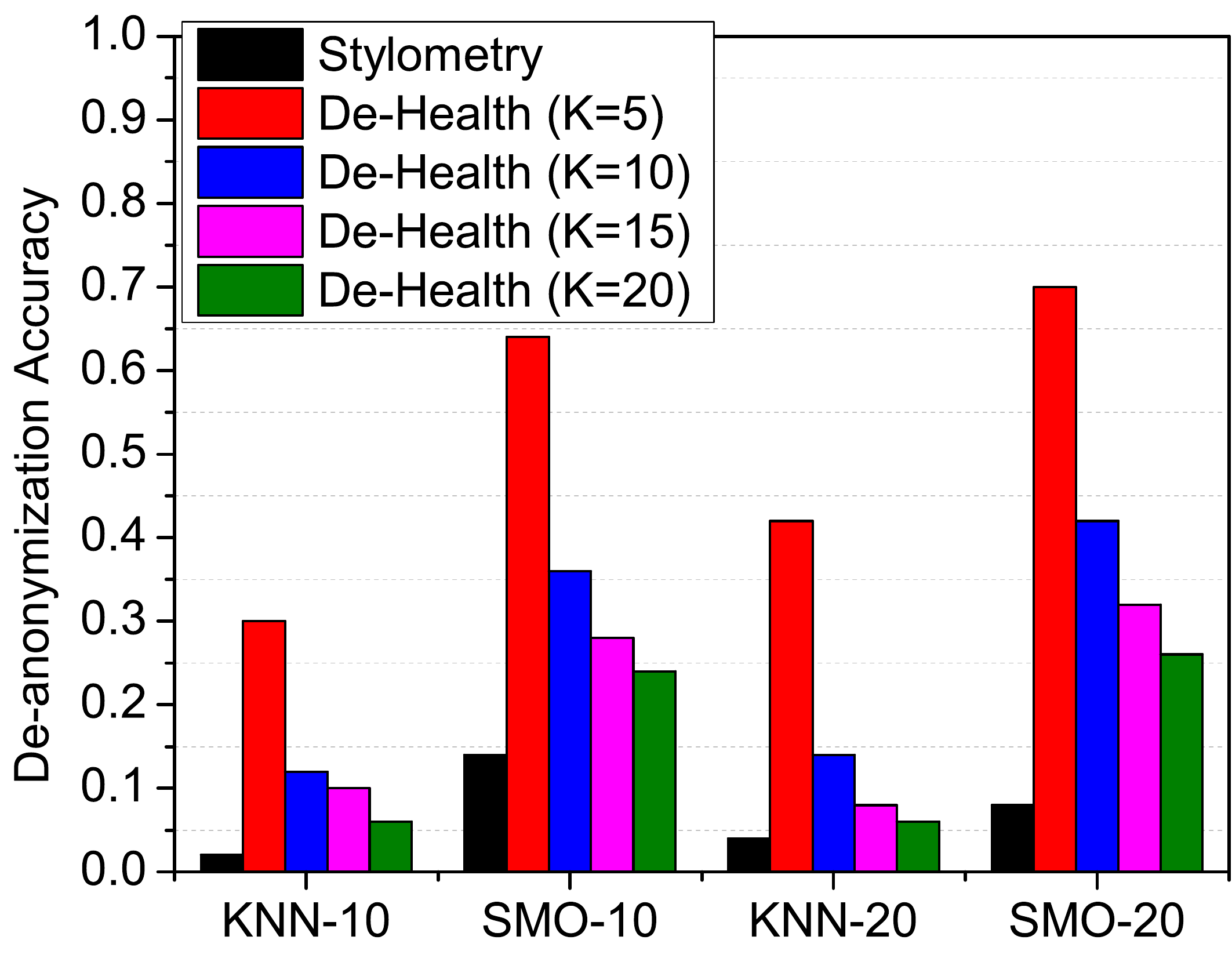}
\caption{DA accuracy (closed-world).} \label{f_dac}
\end{figure}

\textbf{Results.}
Let $Y$ be the number of anonymized users that have true mappings in $\Delta_2$
and $Y_c$ be the number of anonymized users that have true mappings in $\Delta_2$
and are successfully de-anonymized by algorithm $\mathcal{A}$.
Then, the \emph{accuracy} of $\mathcal{A}$ is defined as $Y_c/Y$.

We demonstrate the DA accuracy of De-Health
and Stylometry in Fig.\ref{f_dac}, where $K = 5, 10, 15, 50$
indicate the setting of Top-$K$ DA in De-Health,
and `-10' (e.g., SMO-10) and `-20' (e.g., SMO-20) represent the
evaluation settings with 10 and 20 posts of each user for training/testing, respectively.
From the results, SMO has a better performance than KNN
with respect to de-anonymizing the employed WebMD datasets.

De-Health significantly outperforms Stylometry,
e.g., in the setting of SMO-20, De-Health ($K = 5$)
successfully de-anonyimzes $70\%$ users (with accuracy of $70\%$)
while Stylometry only successfully de-anonymizes $8\%$ users:
($i$) for Stylometry, given 20 (resp., 10) posts
and the average length of WebMD posts is 127.59,
the training data is 2551.8 (resp., 1275.9) words on average,
which might be insufficient for training an effective classifier
to de-anonymize an anonymized user;
and ($ii$) as expected, this demonstrates that
De-Health's Top-$K$ DA phase is very effective,
which can clearly reduce the DA space (from 50 to 5)
with a satisfying successful Top-$K$ DA rate
(consistent with the results in the Top-$K$ DA evaluation).

Interestingly, De-Health has better accuracy for a smaller $K$
than for a larger $K$.
Although a large $K$ implies a high successful Top-$K$
DA rate, it cannot guarantee a better refined (precise)
DA accuracy in the second phase, especially
when the training data for the second phase
(same to Stylometry) are insufficient.
On the other hand, a smaller $K$ is more likely to induce a better
DA performance since it reduces more of the possible DA space.
Therefore,
\emph{when less data are available for training,
the Top-$K$ DA phase is more likely to dominate the
overall DA performance}.

\subsection{Open-world DA}

Now, we evaluate De-Health in a more
challenging setting where the anonymized user may or may not
have a true mapping in the auxiliary data,
i.e., open-world DA.

\subsubsection{Top-$K$ DA}

We start the open-world evaluation from examining
the effectiveness of the Top-$K$ DA of De-Health.

\textbf{Methodology and Settings.}
Leveraging the WebMD and HB datasets,
we construct three open-world DA scenarios under which
the anonymized data and the auxiliary data have the same number of
users and their overlapping user ratios are $50\%$, $70\%$, and $90\%$,
respectively\footnote{Let $n$ be the number of users in WebMD/HB, and $x$ and $y$
be the number of overlapping and non-overlapping users in the auxiliary/anonymized dataset.
Then, it is straightforward to determine $x$ and $y$ by solving the equations:
$x + 2y = n$ and $\frac{x}{x + y} = 50\%$ (resp., $70\%$ and $90\%$).}.
Then, we employ De-Health to examine the Top-$K$ DA
performance in each scenario with the default setting:
for each overlapping user, take half of its data (posts) for training
and the other half for testing;
$c_1 = 0.05$, $c_2 = 0.05$, and $c_3 = 0.9$ (for the same reason as explained before);
$\hbar = 50$; and for the Top-$K$ candidate approach, employ \emph{direct selection}.
All the evaluations are repeated 10 times.
The results are the average of those 10 runs.


\begin{figure}
\centering
\includegraphics[width=3.5in]{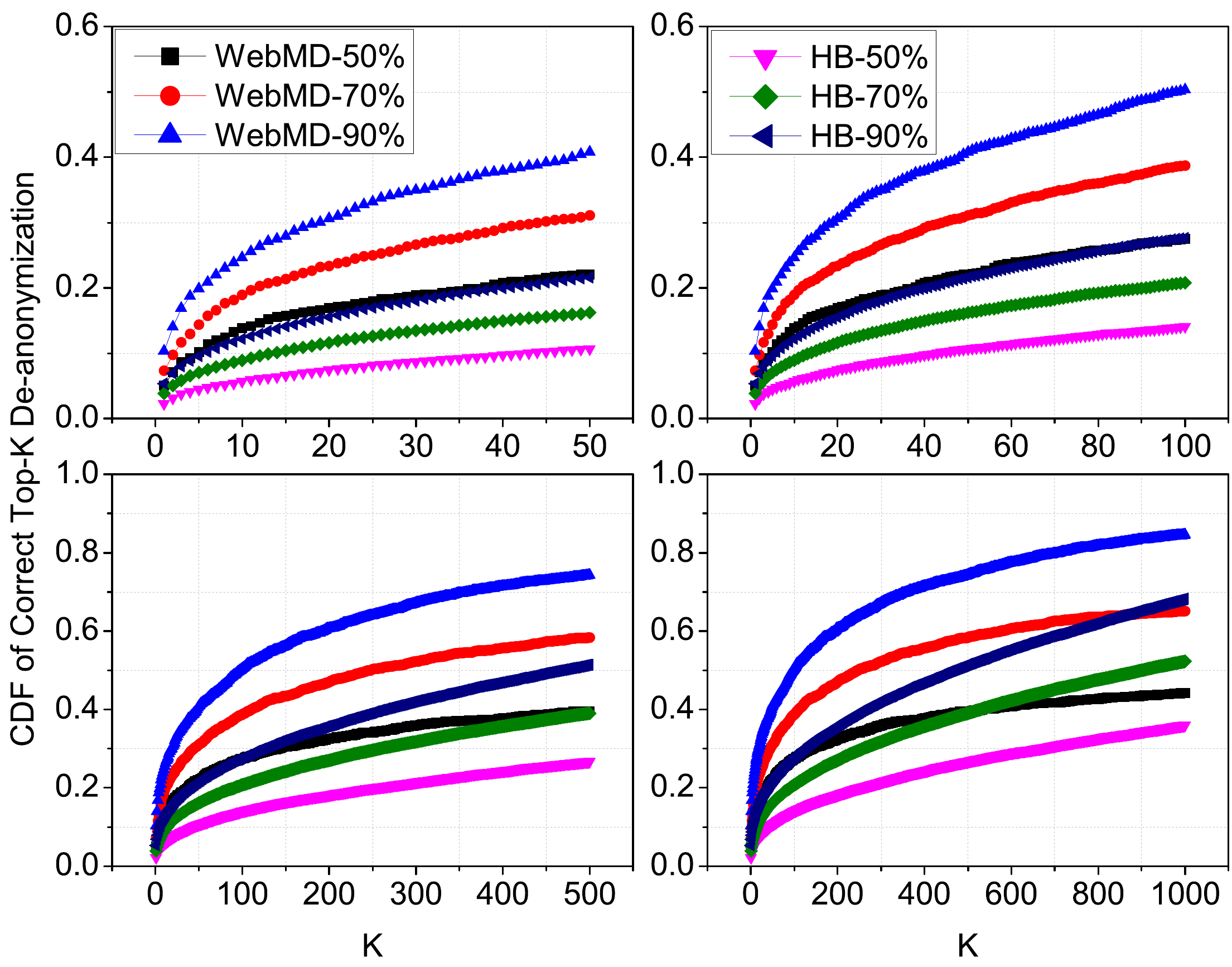}
\caption{CDF of correct Top-$K$ DA (open-world).} \label{f_topko}
\end{figure}

\textbf{Results.}
We show the Top-$K$ DA performance given different
$K$ ranges ($K \in [1, 50]$, $K \in [1, 100]$, $K \in [1, 500]$, and $K \in [1, 1000]$)
in Fig.\ref{f_topko}.
First, similar to that in the closed-world setting,
the CDF of successful Top-$K$ DA increases
with the increasing of $K$ since the true mapping of an anonymized user (if it has)
is more likely to be included in its Top-$K$ candidate set for a large $K$.
Second, De-Health has a better Top-$K$ DA performance
when more users are shared between the anonymized data (graph)
and the auxiliary data (graph). For instance,
when de-anonyming WebMD, the successful Top-500 DA rate is
$58.4\%$ when the overlapping user ratio is $70\%$, while $74.2\%$
when the overlapping user ratio is $90\%$.
This is because a higher overlapping user ratio implies more common users
between the anonymized and auxiliary data, followed by
higher structural similarity between the anonymized and auxiliary UDA graphs.
Thus, De-Health can find the correct Top-$K$ candidate sets for more users
(which are determined by the users' structural similarities).
Third, when comparing closed-world (Fig.\ref{f_topkc})
and open-world (Fig.\ref{f_topko}) Top-$K$ DA,
better performance can be achieved in the closed-world setting.
The reason is the same as our analysis for the second observation.
Finally, under the open-world setting, De-Health can still achieve
a satisfying Top-$K$ DA performance
(compared to the closed-world setting, a larger $K$, e.g., $K=1500$,
might be necessary), and thus significantly reduces the possible
DA space for an anonymized user.

\subsubsection{Refined DA}

Following the Top-$K$ DA, we evaluate the
refined DA performance of De-Health in the open-world setting.
Due to the same reason as analyzed before,
we conduct this group of evaluations on small WebMD datasets,
which is again sufficient to show the performance of De-Health.

\textbf{Methodology and Settings.}
We construct an anonymized dataset and an auxiliary dataset
such that ($i$) each dataset has 100 users and each user has 40 posts;
($ii$) the overlapping user ratio between the two datasets is $50\%$;
and ($iii$) for each overlapping user, half of its posts appear in the
anonymized data while the others appear in the auxiliary data.
Taking the same approach, we construct two other pairs of
anonymized datasets and auxiliary datasets except for with overlapping user ratios
of $70\%$ and $90\%$, respectively.

For De-Health, its default settings are:
$c_1 = 0.05$, $c_2 = 0.05$, and $c_3 = 0.9$;
$\hbar = 5$; $\epsilon = 0.01$ and $l = 10$ for filtering;
the Top-$K$ candidate selection approach is \emph{direct selection};
the leveraged features are the stylometric and structural features
defined in Section \ref{datacollection} and the employed
machine learning techniques are KNN and SMO;
and after classification, we apply for the \emph{mean-verification}
scheme with $r = 0.25$.
We also compare De-Health with \emph{Stylometry} (which can be considered
as equivalent to the second phase of De-Health).
All the experiments are run 10 times and the results are the average
of those 10 runs.


\begin{figure}
 \centering
  \subfigure[DA accuracy]{
    \includegraphics[width=2in]{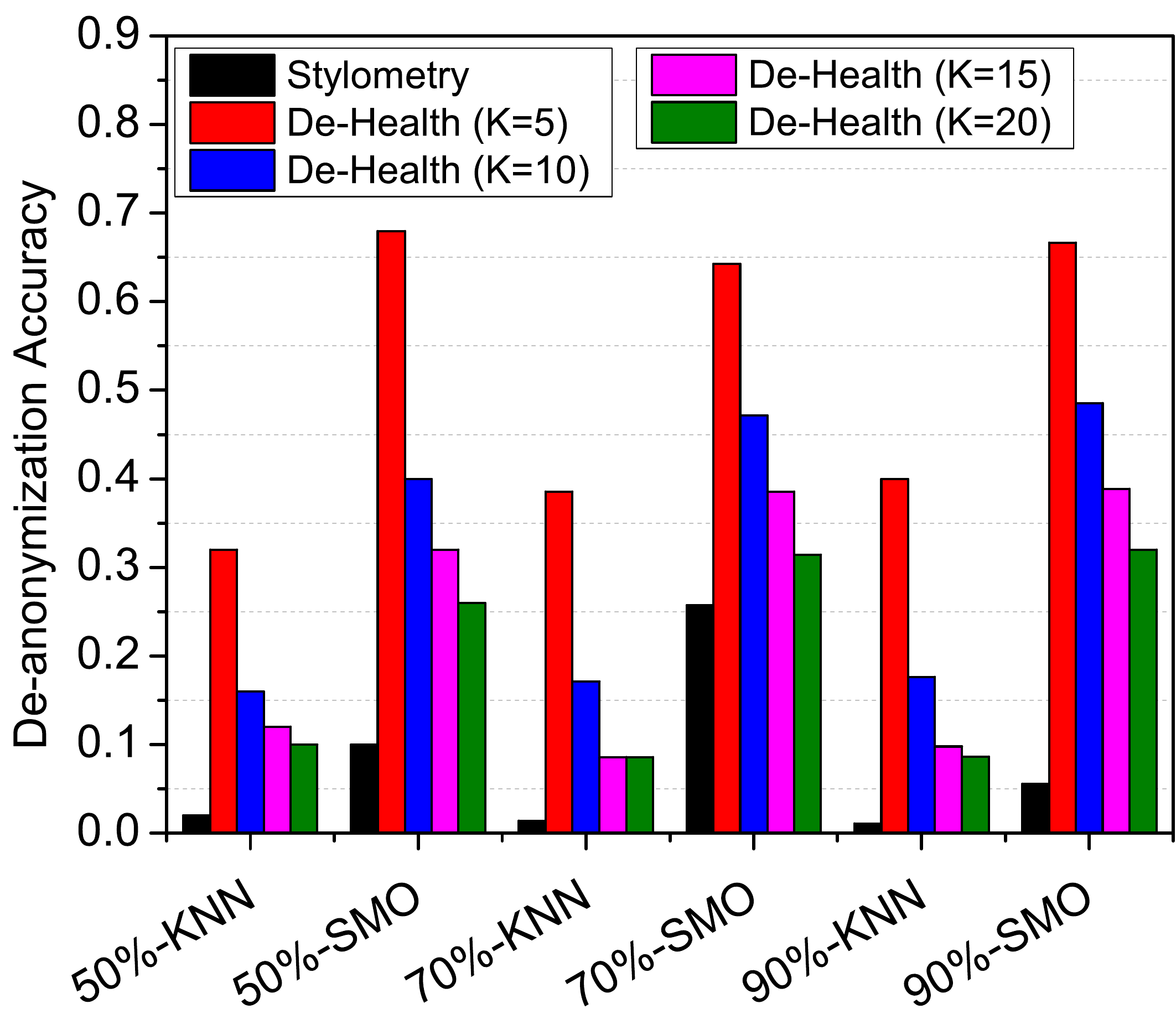}
  }
  \subfigure[FP rate]{
    \includegraphics[width=2in]{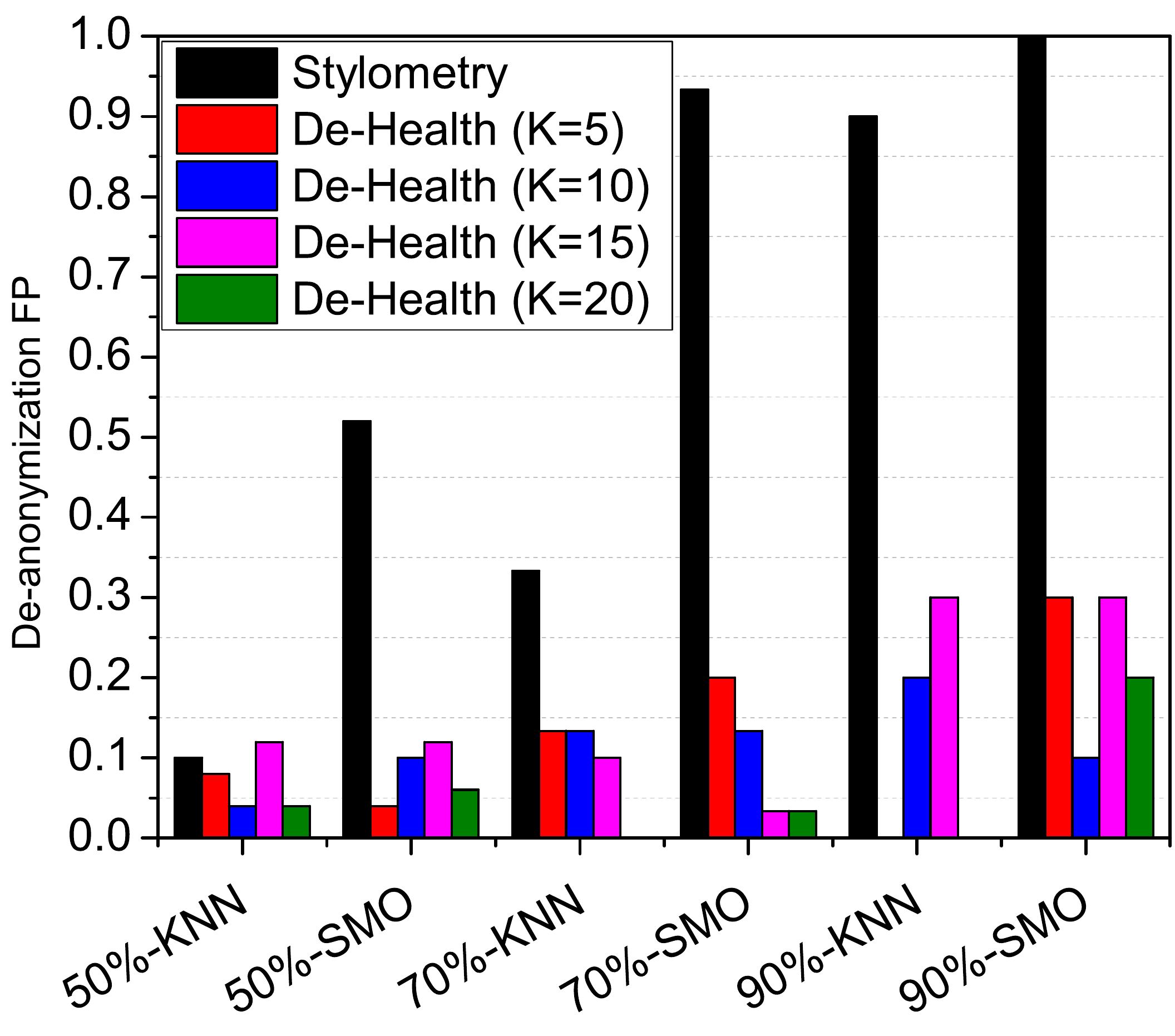}
  }
  \caption{DA accuracy and FP rate (open-world).} \label{f_dao}
\end{figure}

\textbf{Results.}
We report the DA accuracy and False Positive (FP) rate
in Fig.\ref{f_dao}, where $50\%$, $70\%$, and $90\%$ indicate the overlapping user ratios.
First, in the open-world setting, De-Health again significantly
outperforms Stylometry with respect to both DA accuracy
and the FP rate. For instance, in the setting of $50\%$-SMO,
The DA accuracy of De-Health ($K = 5$) is $68\%$
and of Stylometry is $10\%$, respectively; and meanwhile, the FP rate of
De-Health ($K = 5$) is $4\%$ and of Stylometry is $52\%$, respectively.
For Stylometry, insufficient training data is one reason for its poor performance.
In addition, in the open-world DA setting,
non-overlapping users, which can be considered as noise, further degrade its
performance.
On the other hand, for De-Health, there are also two reasons
responsible for its better performance: ($i$) the Top-$K$ DA
reduces the possible DA space while preserving a relatively
high success rate, and thus high DA accuracy is achieved;
and ($ii$) the \emph{mean-verification} scheme eliminates FP DAs
and thus reduces the FP rate.
Second, similar to the closed-world scenario,
De-Health with a smaller $K$ has better DA accuracy (not necessary the FP rate)
than that with a larger $K$. The reason is the same as discussed before:
when less data are available for training in the second phase,
the Top-$K$ DA is more likely to dominate the
overall DA performance of De-Health.
From the figure, we also observe that SMO-trained classifier
induces better performance than KNN-trained classifier in most cases.

\section{Real Identity Identification} \label{linkage}

Leveraging De-Health,
an adversary can now have the medical/health information
of online health services users, e.g., users of
WebMD and HB. On top of the DA
results of De-Health, we present a
\emph{linkage attack} framework to
\emph{link those medical/health information of the service users to real world people}
in this section.

\subsection{Linkage Attack Framework}

In the designed linkage attack framework, we mainly conduct
\emph{username-based linkage} and \emph{avatar-based linkage}.

\textbf{Username-based Linkage.}
For most online health services, the users' usernames are publicly available.
In addition to that, there are many other social attributes that might be publicly available,
e.g., gender, join date, and location of users are available on HB.
In \cite{percaspets11}, Perito et al.
empirically demonstrated that Internet users tend to choose a small
number of correlated usernames and use them across many online services.
They also developed a model to characterize the \emph{entropy} of a given Internet username
and demonstrated that a username with high (resp., low) entropy
is very unlikely (resp., likely) picked by multiple users.
Motivated by this fact, we implement a tool, named \emph{NameLink}, to
semi-automatically connect usernames on one online health service
and other Internet services, e.g., Twitter.

NameLink works in the following manner:
($i$) collect the usernames of the users of an online health service;
($ii$) compute the entropy of the usernames using the technique in \cite{percaspets11}
and sort them in the entropy decreasing order;
($iii$) perform \emph{general} and/or \emph{targeting} online search using the
sorted usernames (leveraging Selenium, which automates browsers  and
imitates user's mouse click, drag, scroll and many other input events).
For general online searches, NameLink searches a username with/without
other attributes (e.g., location) directly, e.g., ``jwolf6589 + California";
for targeted searches, in addition to terms used in general search,
NameLink adds a targeting Internet service, e.g., ``jwolf6589 + Twitter";
and ($iv$) after obtaining the search results, NameLink filters unrelated results
based on predefined heuristics.
The main functionalities of NameLink include:
($i$) \emph{information aggregation}; For instance, there is not too much
information associated with WebMD users. However, there is rich information associated with
HB users (e.g., location) and BoneSmart users (e.g., ages) \cite{bonesmart}.
By linking the users on those three services, we may obtain richer information of WebMD users;
($ii$) \emph{real people linkage}; For instance, for the WebMD users that have high entropy,
e.g., ``jwolf6589", we may try to link them to social network services,
e.g., Twitter, and thus reveal their true identities;
and ($iii$) \emph{cross-validation}. For each user, we may link her
to a real world person using multiple techniques, e.g., the username-based linkage
and the following avatar-based linkage.
Therefore, using the linkage results from different techniques
can further enrich the obtained information as well as
cross-validate the search results, and thus improve the linkage accuracy.

\textbf{Avatar-based Linkage.}
Many online health services, e.g., WebMD, allow users to choose their own avatars.
Thus, many users take this option by uploading an avatar without  awareness
of the privacy implications of their actions.
However, as shown in \cite{ilipolccs15}, those photos may also cause serious privacy
leakage. The reason behind is that a significant amount of users
upload the same photo/avatar across different Internet services (websites).
Similar to NameLink, we develop another semi-automatic tool,
named \emph{AvatarLink}, to link the users of one online health service
to other Internet services, e.g., Facebook, Twitter.
AvatarLink generally follows the same working procedure as
NameLink except for the search engine, which takes either an image URL or user
uploaded image file as a search key.
AvatarLink can also fulfill the same functionalities as NameLink,
i.e., information aggregation, real people linkage, and cross-validation.

\subsection{Evaluation}

We validate the linkage attack framework using the collected WebMD dataset
since all its users have publicly available usernames
and many of them have publicly available avatars.
Note that, \emph{the employed WebMD dataset is collected from a real world
online health service} (and thus \emph{generated by real people}).
Therefore, \emph{it might be illegal, at least improper, to employ NameLink
and AvatarLink to conduct a large-scale linkage attack although we can do that}.
\emph{When linking the medical/health information to real world people,
we only show a proof-of-concept attack and results}.

\textbf{Objectives and Settings.}
Considering that there is not too much information associated with
WebMD users, we have two objectives for our evaluation:
($i$) information aggregation, i.e., enrich the information of
WebMD users; and ($ii$) link WebMD users to real world people,
reveal their identities, and thus compromise their medical/health privacy.

To achieve the first objective, we employ NameLink for
targeting linkage and the targeting service is HB,
which has rich user information. Since we have both a WebMD dataset
and a HB dataset, we limit our linkage to the users within
the available datasets and thus we can do the linkage offline.
Note that, this is a proof-of-concept attack and it can be
extended to large-scale directly.

To achieve the second objective, we employ AvatarLink to link
WebMD users to some well known social network services, e.g., Facebook, Twitter, and LinkedIn.
There are 89,393 users in the WebMD dataset, which are too many
for a proof-of-concept linkage attack.
Thus, we filter avatars (i.e., users) according to four conditions:
($i$) exclude default avatars;
($ii$) exclude avatars depicting non-human objects, such as animals, natural scenes, and logos;
($iii$) exclude avatars depicting fictitious persons;
and ($iv$) exclude avatars with only kids in the picture.
Consequently, we have 2805 avatars left.
When using AvatarLink to perform the linkage attack,
the employed search engine is Google Reverse Image Search.
In order to avoid the violation of Google's privacy and security policies,
we spread the searching task of the 2805 avatars in five days
(561 avatars/day, on average)
and the time interval between two continuous searches is at least
1 minute.

\textbf{Results and Findings.}
For understanding and analyzing the results returned by NameLink and AvatarLink, a challenging task
is to validate their accuracy.
To guarantee the preciseness as much as possible, we manually
validate all the results and only preserve the ones with high confidence.
Specifically, for the results returned by NameLink, in addition to using
the technique in \cite{percaspets11} to filter out results with low entropy usernames,
we manually compare the users' posts on two websites with respect to
writing style and semantics, as well as the users' activity pattern,
e.g., post written time. Interestingly, many linked users post the same description
of their medical conditions on both websites to seek suggestions.
For the results returned by AvatarLink, we manually compare the
person in the avatar and the person in the found picture, and only
results in which we are confident are preserved.

Finally, using NameLink, we successfully link 1676 WebMD users to
HB users and thus, those users' medical records and other
associated information can be combined to provide us (or adversaries)
more complete knowledge about them.
Using AvatarLink, we successfully link 347 WebMD users to
real world people through well known social network services
(e.g., Facebook, Twitter, LinkedIn,
and Google+), which consists $12.4\%$ of the 2805 target users.
Among the 347 WebMD users, more than $33.4\%$
can be linked to two or more social network services,
and leveraging the Whitepage service \cite{whitepage},
detailed social profiles of most users can be obtained.
More interestingly, the WebMD users linked to HB
and the WebMD users linked to real people have 137 overlapping users.
This implies that information aggregation and linkage attacks
are powerful in compromising online health service users' privacy.
Overall, we can acquire most of the 347 users'
full name, medical/health information, birthdate,
phone numbers, addresses, jobs, relatives, friends, co-workers, etc.
Thus, those users' privacy suffers from a serious threat.
For example, after observing the medical/health records of some users,
we can find their sexual orientation, relationships, and related
infectious diseases. More concerning, some of the users even
have serious mental/psychological problems and show suicidal tendency.

\section{Discussion} \label{discussion}


\textbf{De-Health: Novelty versus Limitation.}
As shown in the experiments (Section \ref{experiment}), the Top-$K$
DA of De-Health is effective in reducing the DA
space (from 100K-order of possible space to 100-order of possible space)
while preserving a satisfying precision (having the true mapping of
an anonymized user included into the candidate set).
Further, when the training data for constructing a powerful classifier
are insufficient, such DA space reduction is more helpful
for De-Health to achieve a promising DA accuracy.
Therefore, the Top-$K$ DA is stable and robust.
For the refined DA phase, technically, it can be implemented
by existing benchmark machine learning techniques.
Nevertheless, due to the benefit of the Top-$K$ DA phase,
the possible DA space is reduced by several orders of magnitude,
which enables us to build an effective classifier even with insufficient training data.
Therefore, the Top-$K$ DA together with
the refined DA lead to promising performance
of De-Health in both closed-world and open-world scenarios.

It is important to note that we do not apply advanced anonymization techniques
to the health data when evaluating the performance of De-Health.
This is mainly because no feasible or dedicated anonymization technique
is available for large-scale online health data, to the best of our
knowledge. Actually, developing proper anonymization techniques for
large-scale online health data is a challenging open problem.
The challenges come from ($i$) the data volume is very big,
e.g., WebMD has millions of users that generate millions to billions of
health/medical posts every month;
($ii$) unlike well-structured traditional medical records,
the online health data are generated by millions of different users.
It is a challenging task to organize those unstructured (complex) data;
and ($iii$) different from other kinds of data, health/medical data
have sensitive and important information.
A proper health data anonymization scheme should appropriately
preserve the  data's utility (e.g., preserve the accurate description
of a disease).
We take developing effective online health data anonymization techniques
as a future work.

\textbf{Re-identifiability Analysis: Generic versus Loose.}
In our theoretical analysis of online health data DA,
we quantify the impacts of different data features,
including local interactivity features, global interactivity features,
associated attributes, and stylometric features,
on the anonymity of the data.
We also derive the conditions and probabilities for
successfully de-anonymizing one user or a group of users
in both Top-$K$ and accurate DA.
However, to guarantee the maximum generality, we do not specify
the exact distribution of considered features.
In reality, it is possible to obtain tighter conditions and
probability bounds when specifying the distributions of features
(evidently, this is at the cost of sacrificing the generality
of the theoretical analysis),
e.g., assuming the local interactivity features follow a Poisson distribution.
Therefore, studying the characteristics of different features
and deriving the analysis under some specific distribution
will be another future work.

\textbf{Online Health Data Privacy and Policies.}
Based on our analysis and experimental results (especially the results
of the linkage attack), online health data privacy suffers from serious threats.
 Unfortunately, there is no effective solution
for protecting the privacy of online health service users
from either the technical  perspective or the policy perspective.
Therefore, our results in this paper are expected to shed
light in two areas: ($i$) for our De-Health and linkage attack frameworks
and evaluation results, they are expected to show users, data owners, researchers, and policy makers
the concrete attacks and the corresponding serious privacy leakage;
and ($ii$) for our theoretical analysis,
it is expected to provide researchers and policy makers a clear
understanding of the impacts that different features have on the
data anonymity, and thus help facilitate them to develop
effective online health data anonymization techniques
and proper privacy  policies.

\section{Related Work} \label{related}


\textbf{Hospital/Structured Data Anonymization and DA.}
To anonymize the claims data used in the Heritage Health Prize (HHP)
competition  and ensure that they meet the
HIPAA Privacy
Rule, Emam proposed several anonymization methods based on a risk threshold \cite{emaarbjmir12}.
Fernandes et al. developed an anonymous psychiatric case register,
the Clinical Record Interactive Search (CRIS), based on the EHRs
generated from the South London and Maudsley NHS Trust (SLaM) \cite{ferclobmc13}.
For the scenario of statistical health information release,
Gardner et al. developed SHARE, which can release the information
in a differentially private manner \cite{garxioamia13}.
To defend against the re-assembly attack, Sharma et al. proposed
DAPriv, an encryption-based decentralized architecture for protecting
the privacy of medical data \cite{shasub14}.
In \cite{emajonplos11}, Emam et al. systematically evaluated existing
DA attacks to structured health data.
A comprehensive survey on existing privacy-preserving structured health
data publishing techniques (45+) was given
by Gkoulalas-Divanis et al. in \cite{gkoloujbi14}.

\textbf{Online Health Data.}
In \cite{niezhatkde14}, Nie et al. sought to bridge the vocabulary
gap between health seekers and online healthcare knowledge.
Another similar effort is \cite{luotanir08}, where Luo and Tang
developed iMed, an intelligent medical Web search engine.
Along the line of analyzing users' behavior in searching,
Cartright et al. studied the intentions and attention in exploratory
health search \cite{carwhiir11} and White and Horvitz studied the onset
and persistence of medical concerns in search logs \cite{whihorir12}.
Nie et al. studied automatic disease inference in \cite{niewantkde15}.

\textbf{Health Data Policy.}
In \cite{bar12},
Barth-Jones re-examined the `re-identification' attack of Governor William Weld's
medical information.
In \cite{senferjmir}, Se\={n}or et al. conducted a review of free
web-accessible Personal Health Record (PHR) privacy policies.
In \cite{mcgamia13}, McGraw summarized concerns with the anonymization
standard and methodologies under the HIPAA regulations.
In \cite{hribloamia14}, Hripcsak et al. summarized the ongoing gaps
and challenges of health data use, stewardship, and governance,
along with policy suggestions.
In \cite{emarodbmj15}, Emam et al. analyzed the key concepts
and principles when anonymizing health data while ensuring
it preserves the utility for meaningful analysis.

\textbf{Stylometric Approaches.}
Stylometric techniques have been widely used for compromising the anonymity
of online users. In \cite{abbchetis08}, Abbasi and Chen proposed the use
of stylometric analysis techniques to identify authors based on writing style.
In \cite{kopschlre11}, Koppel et al. studied the authorship attribution problem
in the wild.
Later, in \cite{narpassp12}, Narayanan et al. studied the feasibility of Internet-scale
author identification.
Considering that the closed-world setting does not hold in many real world applications,
Stolerman et al. presented a Classify-Verify framework for open-world
author identification, which performs better in adversarial settings than
traditional author classification.
In \cite{afrbresp12}, Afroz et al. studied the performance of stylometric techniques
when faced with authors who intentionally obfuscate their writing style
or attempt to imitate that of other authors.
In \cite{afrcalsp14}, Afroz et al. investigated stylometry-based adapting
authorship attribution to underground forums and proposed a general multiple
author detection algorithm.
Another scenario of applying stylometric techniques is \cite{calharusenix15},
where Caliskan-Islam et al.de-anonymized programmers via code stylometry.
To defend against stylometry-based author attribution,
McDonald et al. presented Anonymouth \cite{mcdafrpets12}.
In \cite{breafrtissec12}, Brennan et al. proposed a framework for creating
adversarial passages, which includes obfuscation, imitation, and translation
techniques.

%
%

%

\section{Conclusion} \label{conclusion}

In this paper, we study the privacy of online health data.
Our main conclusions are four-fold.
First, we present a novel two-phase online health data DA attack,
named De-Health,
which can be applied to both closed-world and open-world
DA settings.
Second, we conduct the first theoretical analysis on the soundness
and effectiveness of online health data DA.
Our analysis explicitly shows the conditions and probabilities
of successfully de-anonymizing one user or a group of users
in both exact DA and Top-$K$ DA.
Third, leveraging two large real world online health datasets,
we validate the performance of De-Health.
De-Health can significantly reduce the DA space
while preserving high accuracy.
Even in the scenario where training data are insufficient,
De-Health still achieves promising DA accuracy.
Finally, we present a linkage attack framework that can link
online health data to real world people and thus clearly
demonstrate the vulnerability of existing online health data.
Our findings have meaningful implications to researchers
and policy makers in helping them understand the privacy
vulnerability of online health data and develop effective anonymization techniques
and proper privacy policies.

%
%

\appendix

\subsection{Example Posts from WebMD and HB} \label{posts}

Below are example posts from WebMD and HB, respectively.
\begin{quote}
\emph{WebMD User ***:}
Hi I have hep c genotype 3b I guess it's a rare strain I'm 29 my viral load has gone from 10thousand to 3 million in 5 months and it's jumped to 8 million in the last month what's going on? I'm in the process of screening for treatment but going up 5millions in one month? Is that possible I am extremely extremely I'll. I'm also battling with tapering off of ativan and methadone my doctor says that these meds don't activate hep c or aggravate I mean. My alt is like 400 and last like 200 has anyone experience this? I'm in severe withdrawal even not tapering just holding on my meds but I'm so sick and scared I've tried to detox off my meds so many times that I have what is called kindling. My nervous system is completely destabilized and I have psychotic trauma symptoms from the repeated attempts it's going to take years to taper off my meds but can someone give me the lower down on all this hep c and viral load stuff my doctors are blown away how fast it is progressing

\emph{HB User ***:}
hello everyone, I hope I am posting in the right place
I have been sexually active for 2 and a half years, and only had 2 partners. I am now with my boyfriend for two years.
I was in Germany for one month and a half, and when I came back we had unprotected sex 2 days in a row.
That week I started to have some abdominal pain when I peed. It felt like my left side was being crushed. A few days after that, when I went to pee, the whole area, both left and right side hurt, and the pain kept being there for another 10-15 minutes.
Anyway, I went and got tests done, I have someone who works in a lab and I got all the tests done, and I was diagnosed with mycoplasma huminis and ureaplasma urealyticum infections, having to take doxicilin, suppositories and ovules(?) (sorry, English is not my first language).
Now my dilemma is this:
I have been told it is an STD. How can it be an STD when I was my boyfriend's first and we've been together for 2 years without any kind of these problems and have had unprotected sex before? What are the causes of this?
I'm really confused and don't know what to think.
\end{quote}

\begin{figure}
\centering
\includegraphics[width=2.7in]{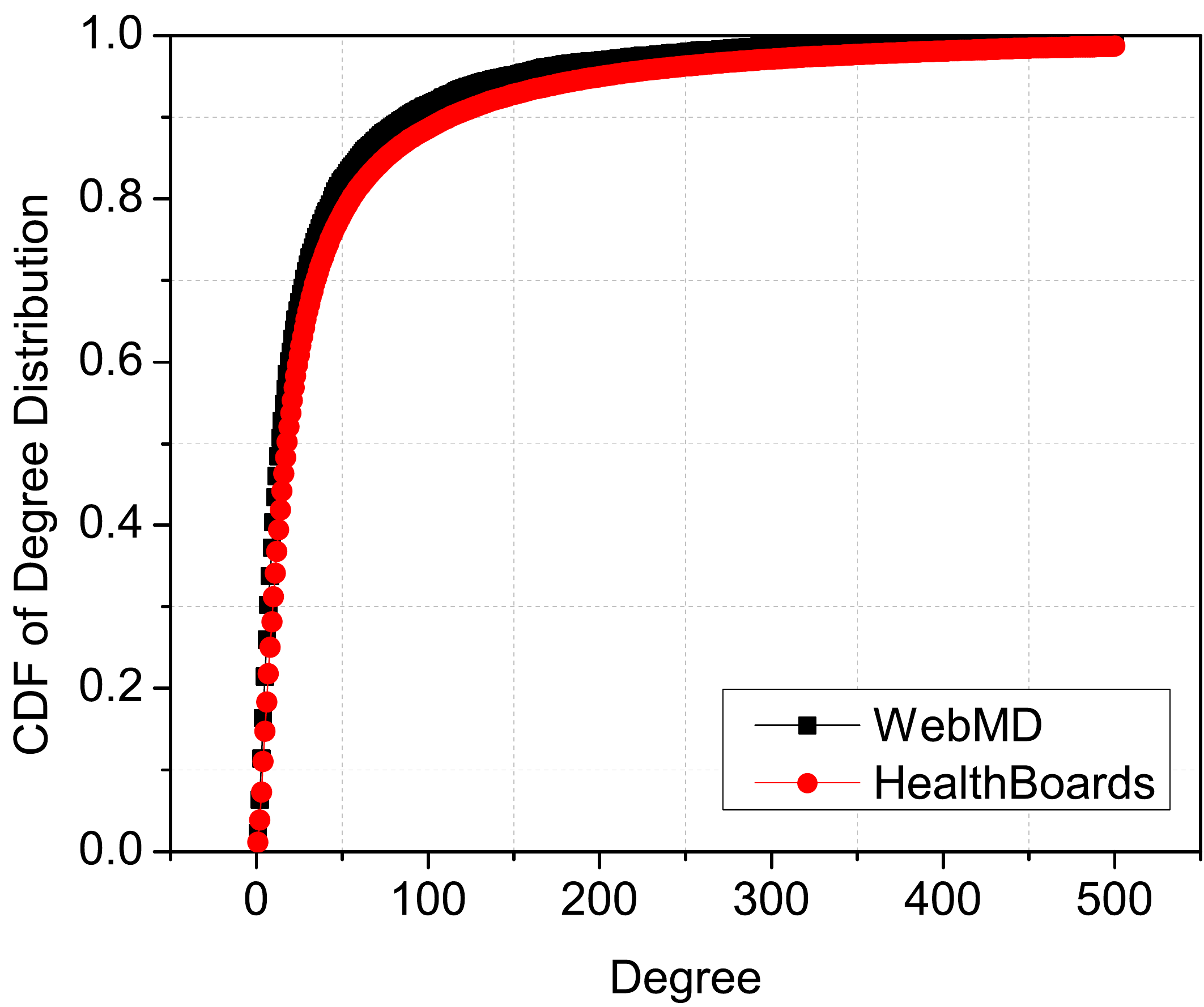}
\caption{User degree distribution.} \label{f_degdis}
\end{figure}

\begin{figure}
\centering
\includegraphics[width=2.8in]{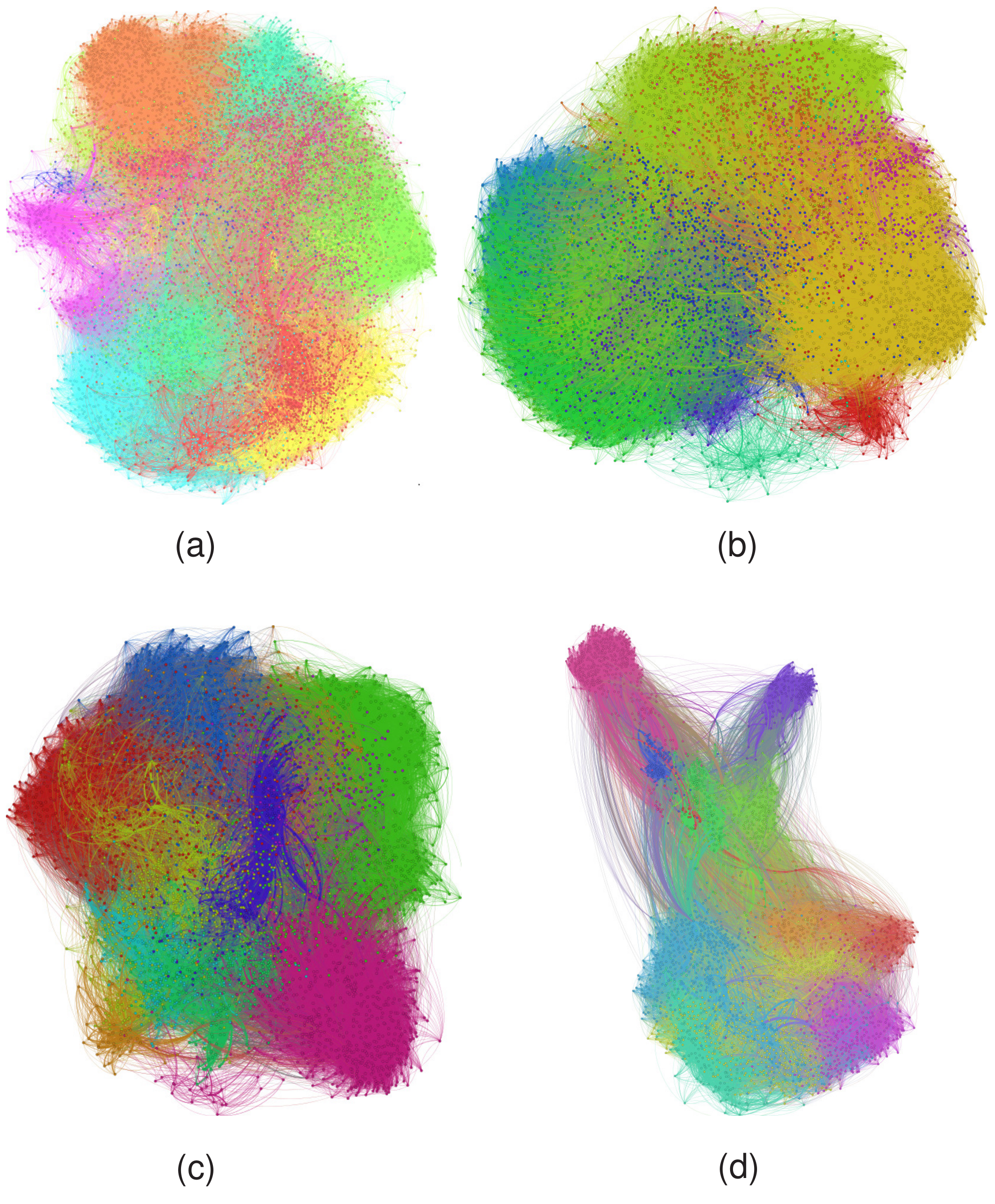}
\caption{WebMD graph community structure.} \label{f_community}
\end{figure}

\subsection{Degree Distribution and Community Structure
of the Correlation Graph} \label{degcom}

We demonstrate the degree distributions of the WebMD graph
and the HB graph in Fig.\ref{f_degdis}.
Basically, the degree of most of the users in the two datasets are low.
Employing the WebMD graph as an example, we show its community structure
under different settings in Fig.\ref{f_community}
(depicted using the tool in \cite{communitytool}),
where (a) shows the community structure of the original WebMD graph,
and (b), (c), and (d) show the community structures of the WebMD graph
by omitting the users with degree less than 11, 21, and 31, respectively.
In each scenario, the graph is not connected (consisting of several components)
and about 10 -- 100 communities can be identified, which implies that
the graph's connectivity is not strong.

\subsection{Proofs} \label{proof}

\textbf{Proof of Theorem \ref{t1}.}
Let $X = \frac{\lambda + \overline{\lambda}}{2}$.
We prove this theorem by considering the following two cases.
When $\lambda < \overline{\lambda}$, we have
$
\Pr(f(u, u') \geq f(u, v))
 \leq \Pr(f(u, u') \geq X) + \Pr(f(u, v) \leq X)
 = \Pr(f(u, u') \geq \frac{\lambda + \overline{\lambda}}{2})
+ \Pr(f(u, v) \leq \frac{\lambda + \overline{\lambda}}{2}).
$
Let $\varepsilon_1 = \frac{\overline{\lambda} - \lambda}{2\lambda}$ and
$\varepsilon_2 = \frac{\overline{\lambda} - \lambda}{2\overline{\lambda}}$.
We have
$
\Pr(f(u, u') \geq f(u, v))
 \leq \Pr(f(u, u') \geq (1 + \varepsilon_1) \lambda)
+ \Pr(f(u, v) \leq (1 - \varepsilon_2)\overline{\lambda}).
$
Applying the generalized Chernoff bound \cite{chernoff}, we have
$
\Pr(f(u, u') \geq f(u, v))
 \leq \exp(-\frac{2\varepsilon_1^2 \lambda^2}{\theta^2})
 + \exp(-\frac{\varepsilon_2^2 \overline{\lambda}^2}{\overline{\theta}^2})
 = \exp(-\frac{2\lambda^2}{\theta^2} \frac{(\overline{\lambda} - \lambda)^2}{4 \lambda^2})
+ \exp(-\frac{\overline{\lambda}^2}{\overline{\theta}^2}
\frac{(\overline{\lambda} - \lambda)^2}{4 \overline{\lambda}^2})
 = \exp(-\frac{(\overline{\lambda} - \lambda)^2}{2\theta^2})
+ \exp(-\frac{(\overline{\lambda} - \lambda)^2}{4\overline{\theta}^2})
 \leq \exp(-\frac{(\overline{\lambda} - \lambda)^2}{4\theta^2})
+ \exp(-\frac{(\overline{\lambda} - \lambda)^2}{4\overline{\theta}^2})
 \leq 2\exp(-\frac{(\lambda - \overline{\lambda})^2} {4 \delta^2}).
$
Then, we design a $\mathcal{M}$ under which $u \rightarrow
\arg\min\limits_{x \in \{u', v\}} f(u, x)$,
i.e., $u$ is de-anonymized to the one in $\{u', v\}$
who generates the smallest $f(u, \cdot)$ value.
Then, according to our above analysis, we have
$\Pr(f(u, u') \geq f(u, v)) \leq 2\exp(-\frac{(\lambda - \overline{\lambda})^2} {4 \delta^2})$,
which implies $\Pr(f(u, u') < f(u, v)) \geq 1- 2\exp(-\frac{(\lambda - \overline{\lambda})^2} {4 \delta^2})$,
i.e., $\Pr(u \stackrel{\{u', v\}}{\rightarrow} u') \geq 1 - 2\exp(-\frac{(\lambda - \overline{\lambda})^2}
{4 \delta^2})$.

Now, we consider the second case that $\lambda > \overline{\lambda}$.
Let $\varepsilon_1 = \frac{\lambda - \overline{\lambda}}{2\lambda}$ and
$\varepsilon_2 = \frac{\lambda - \overline{\lambda}}{2\overline{\lambda}}$.
Then, applying the generalized Chernoff bound, we have
$
\Pr(f(u, u') \leq f(u, v))
 \leq \Pr(f(u, u') \leq X) + \Pr(f(u, v) \geq X)
 = \Pr(f(u, u') \leq \frac{\lambda + \overline{\lambda}}{2})
+ \Pr(f(u, v) \geq \frac{\lambda + \overline{\lambda}}{2})
 = \Pr(f(u, u') \leq (1 - \varepsilon_1) \lambda)
+ \Pr(f(u, v) \geq (1 + \varepsilon_2)\overline{\lambda})
 \leq \exp(-\frac{\varepsilon_1^2 \lambda^2}{\theta^2})
 + \exp(-\frac{2\varepsilon_2^2 \overline{\lambda}^2}{\overline{\theta}^2})
 \leq \exp(-\frac{(\lambda - \overline{\lambda})^2}{4\theta^2})
+ \exp(-\frac{(\lambda - \overline{\lambda})^2}{2\overline{\theta}^2})
 \leq 2\exp(-\frac{(\lambda - \overline{\lambda})^2} {4 \delta^2}).
$
Then, we design a $\mathcal{M}$ under which $u \rightarrow
\arg\max\limits_{x \in \{u', v\}} f(u, x)$.
Then, according to our above analysis, we have
$\Pr(f(u, u') \leq f(u, v)) \leq 2\exp(-\frac{(\lambda - \overline{\lambda})^2} {4 \delta^2})$,
which implies $\Pr(f(u, u') > f(u, v)) \geq 1- 2\exp(-\frac{(\lambda - \overline{\lambda})^2} {4 \delta^2})$,
i.e., $\Pr(u \stackrel{\{u', v\}}{\rightarrow} u') \geq 1 - 2\exp(-\frac{(\lambda - \overline{\lambda})^2}
{4 \delta^2})$.

\textbf{Proof of Corollary \ref{c1}.}
Since $|\lambda - \overline{\lambda}|/2\theta
\geq \sqrt{2\ln n + \ln 2}$, we have
$\frac{(\lambda - \overline{\lambda})^2}{4 \theta^2} \geq 2\ln n + \ln 2$.
Then, according to Theorem \ref{t1}, we have
$\Pr(u \stackrel{\{u', v\}}{\rightarrow} u') \geq 1 - 2\exp(-\frac{(\lambda - \overline{\lambda})^2}
{4 \delta^2}) \geq 1 - 2 \exp(-2\ln n - \ln 2) = 1 - \frac{1}{n^2}$.
Then, as $n \rightarrow \infty$, $\Pr(u \stackrel{\{u', v\}}{\rightarrow} u') \rightarrow 1$
according to the Borel-Cantelli Lemma, i.e., it is a.a.s.
that $u$ can be successfully de-anonymized from $\{u', v\}$.

\textbf{Proof of Corollary \ref{c2}.}
First, we give the design of $\mathcal{M}$, which is similar to
that in Theorem \ref{t1}.
If $\lambda < \overline{\lambda}$, $\mathcal{M}$ is designed as
$\mathcal{M}: u \rightarrow \arg \min\limits_{x \in V_2} f(u, x)$;
otherwise, if $\lambda > \overline{\lambda}$, $\mathcal{M}$ is designed as
$\mathcal{M}: u \rightarrow \arg \max\limits_{x \in V_2} f(u, x)$.
Then, to prove this corollary, it is equivalent to prove that
$\forall v \in V_2$ and $v \neq u'$, it is a.a.s. that
$\mathcal{M}$ can successfully de-anonymize $u$
from $\{u', v\}$, i.e., $\Pr(u \stackrel{\{u', v\}}{\rightarrow} u')
\stackrel{n \rightarrow \infty}{\rightarrow}  1$.
Let $\mathcal{E}$ be the event that $\exists v \in V_2$
such that $\mathcal{M}$ cannot successfully de-anonymize
$u$ from $\{u', v\}$.
Then, according to Boole's inequality and Theorem 1, we have
$
\Pr(\mathcal{E})
 = \Pr(\bigcup\limits_{v \in V_2, v \neq u'} \mathcal{M}
\text{ cannot de-anonymize $u$ from $\{u', v\}$})
 \leq \sum\limits_{v \in V_2, v \neq u'} \Pr(\mathcal{M}
\text{ cannot de-anonymize $u$ from $\{u', v\}$})
 \leq \sum\limits_{v \in V_2, v \neq u'} 2\exp(-\frac{(\lambda - \overline{\lambda})^2} {4 \delta^2}).
$
Considering that $\lambda \neq \overline{\lambda}$ and $|\lambda - \overline{\lambda}|/2\theta
\geq \sqrt{2\ln n + \ln 2n_2}$, we have
$
\Pr(\mathcal{E})
 \leq \sum\limits_{v \in V_2, v \neq u'} 2 \exp(- 2\ln n - \ln 2n_2)
 \leq 2n_2 \exp(- 2\ln n - \ln 2n_2)
 =1/n^2.
$
Then, as $n \rightarrow \infty$, $\Pr(\mathcal{E}) \rightarrow 0$
according to the Borel-Cantelli Lemma, which implies that
$\Pr(u \stackrel{V_2}{\rightarrow} u')
\stackrel{n \rightarrow \infty}{\rightarrow}  1$, i.e., it is a.a.s.
that $u$ can be successfully de-anonymized from $V_2$ by $\mathcal{M}$.

\textbf{Proof of Theorem \ref{t2}.}
We employ the same $\mathcal{M}$ design as in Corollary \ref{c2}
to de-anonymize the users in $V_\alpha$.
Let $\mathcal{E}$ be the event that $\Delta_1$ is $\alpha$-re-identifiable
and $\mathcal{E}^c$ be its complementary event.
Then, according to Boole's inequality and Theorem 1, we have
$
\Pr(\mathcal{E}^c)
 = \bigcup\limits_{u \in V_\alpha} \Pr(\mathcal{M}
$ cannot de-anonymize $u$ from $\{u', v\}$ for some $v \in V_2$ $)
 = \bigcup\limits_{u \in V_\alpha} \bigcup\limits_{v \in V_2, v\neq u'} \Pr(\mathcal{M}
$ cannot de-anonymize $u$ from $\{u', v\}$ $)
 \leq \sum\limits_{u \in V_\alpha} \sum\limits_{v \in V_2, v \neq u'} \Pr(\mathcal{M}
$ cannot de-anonymize $u$ from $\{u', v\}$ $)
 \leq 2\alpha n_1 n_2\exp(-\frac{(\lambda - \overline{\lambda})^2} {4 \delta^2})
 = \exp(\ln 2\alpha n_1 n_2 -\frac{(\lambda - \overline{\lambda})^2} {4 \delta^2}).
$
Then, we have $\Pr(\mathcal{E}) = 1 - \Pr(\mathcal{E}^c)
\geq 1 - \exp(\ln 2\alpha n_1 n_2 -\frac{(\lambda - \overline{\lambda})^2} {4 \delta^2})$.

\textbf{Proof of Corollary \ref{c3}.}
Since $|\lambda - \overline{\lambda}|/2\theta \geq \sqrt{2\ln n + \ln 2 \alpha n_1 n_2}$,
we have $\frac{(\lambda - \overline{\lambda})^2}{4\theta^2} \geq
2\ln n + \ln 2 \alpha n_1 n_2$.
Then, according to Theorem \ref{t2}, we have
$\Pr(\Delta_1$ is $\alpha$-re-identifiable$)
\geq 1 - \exp(\ln 2\alpha n_1 n_2 -\frac{(\lambda - \overline{\lambda})^2} {4 \delta^2})
= 1 -1/n^2$. Then, as $n \rightarrow \infty$, we have
$\Pr(\Delta_1$ is $\alpha$-re-identifiable$)
\rightarrow  1$, i.e., it is a.a.s.
that $\Delta_1$ is $\alpha$-re-identifiable.

\textbf{Proof of Theorem \ref{t3}.}
First, we discuss how to construct $\mathcal{M}$ for $u$.
If $\lambda < \overline{\lambda}$, $\mathcal{M}$
is designed to return $C_u$ as the $K$ auxiliary users such that they have
the Top-$K$ smallest $f(u, \cdot)$.
Otherwise, if $\lambda > \overline{\lambda}$, $\mathcal{M}$
is designed to return $C_u$ as the $K$ auxiliary users such that they have
the Top-$K$ largest $f(u, \cdot)$.

Then, we prove the first conclusion of this theorem.
Towards this objective, it is equivalent to prove that
$\forall v \in V_2 \setminus C_u$, $\mathcal{M}$ can successfully
de-anonymize $u$ from $\{u', v\}$.
Let $\mathcal{E}$ be the event that $\exists v \in V_2 \setminus C_u$
such that $\mathcal{M}$ cannot successfully
de-anonymize $u$ from $\{u', v\}$.
Then, using Theorem \ref{t1}, we have
$
\Pr(\mathcal{E})
 = \Pr(\bigcup\limits_{v \in V_2 \setminus C_u, v\neq u'}
\mathcal{M} $ cannot successfully de-anonymize $u$ from $\{u', v\}$ $)
 \leq \sum\limits_{v \in V_2 \setminus C_u, v\neq u'}\Pr(
\mathcal{M} $ cannot successfully de-anonymize $u$ from $\{u', v\}$ $)
 \leq 2 (n_2 - K) \exp(-\frac{(\lambda - \overline{\lambda})^2} {4 \delta^2})
 = \exp(\ln 2 (n_2 - K) -\frac{(\lambda - \overline{\lambda})^2} {4 \delta^2}).
$
Thus, $\Pr(u \rightarrow C_u) = 1 - \Pr(\mathcal{E}) \geq 1 -
\exp(\ln 2 (n_2 - K) -\frac{(\lambda - \overline{\lambda})^2} {4 \delta^2})$.

Now, we prove the second conclusion.
If $|\lambda - \overline{\lambda}|/2\theta \geq \sqrt{\ln 2(n_2 - K) + 2 \ln n}$,
we have $\frac{(\lambda - \overline{\lambda})^2} {4 \delta^2}
\geq\ln 2(n_2 - K) + 2 \ln n$.
Then, we have $\Pr(\mathcal{E}) \leq 1/n^2$.
According to the Borel-Cantelli Lemma, we have $\Pr(\mathcal{E}) \rightarrow 0$
as $n \rightarrow \infty$, which implies $\Pr(u \rightarrow C_u)
\stackrel{n \rightarrow \infty}{\rightarrow}  1$, i.e., it is a.a.s.
that $u$ is Top-$K$ re-identifiable.

\textbf{Proof of Theorem \ref{t4}.}
In this theorem, we use the same $\mathcal{M}$ as described in
Theorem \ref{t3} to de-anonymize each user in $V_\alpha$.
Now, we prove the first conclusion.
Let $\mathcal{E}$ be the event that $\exists u \in V_\alpha$ and
$\exists v \in V_2 \setminus C_u$ such that $\mathcal{M}$ cannot successfully
de-anonymize $u$ from $\{u', v\}$.
Then,
$
\Pr(\mathcal{E})
 = \bigcup\limits_{u \in V_\alpha} \bigcup\limits_{v \in v \in V_2 \setminus C_u, v\neq u'}
\Pr($$\mathcal{M}$ cannot successfully
de-anonymize $u$ from $\{u', v\}$ $)
 \leq \sum\limits_{u \in V_\alpha} \sum\limits_{v \in v \in V_2 \setminus C_u, v\neq u'}
\Pr($ $\mathcal{M}$ cannot successfully
de-anonymize $u$ from $\{u', v\}$ $)
 \leq 2\alpha n_1 (n_2 - K) \exp(-\frac{(\lambda - \overline{\lambda})^2} {4 \delta^2})
 = \exp(\ln 2\alpha n_1 (n_2 - K) -\frac{(\lambda - \overline{\lambda})^2} {4 \delta^2}).
$
Thus, we have $\Pr(V_\alpha: u \rightarrow C_u) = 1 - \Pr(\mathcal{E})
\geq 1 -
\exp(\ln 2 \alpha n_1 (n_2 - K) -\frac{(\lambda - \overline{\lambda})^2} {4 \delta^2})$.

For the second conclusion, since
$|\lambda - \overline{\lambda}|/2\theta \geq \sqrt{\ln 2\alpha n_1 (n_2 - K) + 2 \ln n}$,
we have $\frac{(\lambda - \overline{\lambda})^2} {4 \delta^2}
\geq \ln 2\alpha n_1 (n_2 - K) + 2 \ln n$. Then, we have $\Pr(\mathcal{E}) = 1/n^2$
and as $n \rightarrow \infty$, $\Pr(\mathcal{E}) \rightarrow 0$ according to
the Borel-Cantelli Lemma, which implies that $\Pr(V_\alpha: u \rightarrow C_u)
\stackrel{n \rightarrow \infty}{\rightarrow}  1$, i.e., it is a.a.s.
that $\Delta_1$ is Top-$K$ $\alpha$-re-identifiable.

\end{document}